\documentclass[ prx, twocolumn, nofootinbib, floatfix]{revtex4-1} 

\usepackage{times}

\usepackage[mathscr]{euscript}
\usepackage{amsmath}
\usepackage{graphicx}
\usepackage{dcolumn}
\usepackage{bm}
\usepackage{epsfig}
\usepackage{amssymb,latexsym,mathrsfs}
\usepackage{graphicx}
\usepackage{color}
\usepackage{hyperref}
\usepackage{float}
\usepackage{diagbox}

\definecolor{vividviolet}{rgb}{0.62, 0.0, 1.0}
\definecolor{amaranth}{rgb}{0.9, 0.17, 0.31}
\definecolor{palatinateblue}{rgb}{0.15, 0.23, 0.89}
\definecolor{brightpink}{rgb}{1.0, 0.0, 0.5}
\definecolor{cornflowerblue}{rgb}{0.39, 0.58, 0.93}
\definecolor{deepcarminepink}{rgb}{0.94, 0.19, 0.22}
\definecolor{radicalred}{rgb}{1.0, 0.21, 0.37}
\definecolor{blueblue}{RGB}{21,47,181}
\definecolor{greengreen}{RGB}{65,166,16}

\hypersetup{ linktoc=all,
    colorlinks, linkcolor={blueblue},
    citecolor={greengreen}, urlcolor={blueblue}
}

\newcommand{\be}{\begin{equation}}
\newcommand{\ee}{\end{equation}}
\newcommand{\bs}{\begin{split}} 
\newcommand{\bea}{\begin{eqnarray}}
\newcommand{\eea}{\end{eqnarray}}

\renewcommand{\d}[1]{\ensuremath{\operatorname{d}\!{#1}}}
\def\nn{\nonumber} 

\usepackage{mathtools}
\newcommand{\D}{\mathrm{d}}
\newsavebox{\myhbar}

\begin{document}

\title{Analog particle production model for general classes of Taub-NUT black holes}

\author{Joshua Foo${}^{1}$}
\email{joshua.foo@uqconnect.edu.au}
\affiliation{Centre for Quantum Computation \& Communication Technology, School of Mathematics \& Physics,
University of Queensland, St.~Lucia, Queensland, 4072, Australia}
\author{Michael R.R. Good${}^{2}$}
\email{michael.good@nu.edu.kz}
\affiliation{Physics Dept. \& Energetic Cosmos Laboratory, Nazarbaev University, 
Nur-Sultan, 010000, Qazaqstan}
\author{Robert B. Mann${}^{3,4}$}
\email{rbmann@uwaterloo.ca}
\affiliation{Department of Physics \& Astronomy, University of Waterloo,
\\Waterloo, Ontario, N2L 3G1, Canada\\
Institute for Quantum Computing, University of Waterloo,
\\Waterloo, Ontario, N2L 3G1, Canada}

\begin{abstract} 
We derive a correspondence between the Hawking radiation spectra emitted from general classes of Taub-NUT black holes with that induced by the relativistic motion of an accelerated Dirichlet boundary condition (i.e.\ a perfectly reflecting mirror) in (1+1)-dimensional flat spacetime. We demonstrate that the particle and energy spectra is thermal at late-times and that particle production is suppressed by the NUT parameter. We also compute the radiation spectrum in the rotating, electrically charged (Kerr-Newman) Taub-NUT scenario, and the extremal case, showing explicitly how these parameters affect the outgoing particle and energy fluxes.
\end{abstract}

\date{\today} 

\maketitle 
\section{Introduction}\label{sec:Intro}
Taub-NUT black holes are a simple yet instructive electrovacuum solution to the Einstein-Maxwell field equations \cite{Taub:1950ez,Newman:1963yy}. The Taub-NUT metric is a generalisation of the Schwarzschild metric with the addition of the so-called NUT parameter, $l$, and has played an important role in our understanding of the AdS/CFT correspondence \cite{Hawking:1998ct,Chamblin:1998pz,Emparan:1999pm,Mann:1999pc}. Its initial discovery by Taub \cite{Taub:1950ez} and the subsequent coordinate extension applied to it by Newman, Unti and Tamburino (NUT) \cite{Newman:1963yy} led to the eventual discovery of the well-known rotating Kerr black hole solution \cite{kerr1963gravitational}. 
\\\\
\indent In this paper, we propose a simple, (1+1)-dimensional model describing the Hawking radiation \cite{Hawking:1974sw} properties of general classes of Lorentzian Taub-NUT black holes, known as the accelerated boundary correspondence (ABC). The model associates the origin of the black hole coordinates in (3+1)-dimensions with the trajectory of an accelerated mirror (i.e.\ a perfectly reflecting boundary) in (1+1)-dimensional Minkowski spacetime \cite{Davies:1976hi,Davies:1977yv,DeWitt:1975ys}. The relativistic trajectory of the mirror, which rapidly changes the boundary conditions of incoming field modes, induces particle production from the quantum vacuum \cite{Birrell:1982ix,Fabbri}. Recently, this model has been applied to the well-known Schwarzschild \cite{Good:2016oey}, Reissner-Nordstr\"om (RN) \cite{good2020particle}, Kerr \cite{Good:2020fjz}, 
and Kerr-Newman metrics \cite{Foo:2020bmv}, where analytic expressions for the spectra and the late-time thermal emission were derived. 
\\\\
\indent We extend upon the aforementioned studies by considering general classes of Taub-NUT black holes, including those with non-zero angular momentum and charge. Motivated by prior insights gleaned regarding the effect of spin and charge on the radiation, here we ask how the presence of the NUT charge affects particle production and energy emission. Our results show that in general, the presence of the NUT charge inhibits particle production for non-extremal black holes. Meanwhile for the extremal (rotating, electrically and NUT charged) case, we find that the particle and energy spectrum is non-monotonic with increasing $l$. The utility of the ABC approach lies in its ability to elicit simple, closed-form expressions for the aforementioned quantities without relying on approximations which have been known to breakdown in certain regimes \cite{Kerner:2006vu}. 
\\\\
\indent Our paper is organised as follows: in Sec.\ \ref{sec:metacc}, we introduce the field-theoretic details for the Taub-NUT metric and the associated mirror trajectory in (1+1)-dimensional Minkowski spacetime. 
We then calculate the energy and particle spectrum of the outgoing radiation, demonstrating its thermal character at late-times. In Sec.\ \ref{KNTN}, we extend our analysis to the rotating, electrically charged version of the Taub-NUT spacetime (Kerr-Newman Taub-NUT, or KNTN). We conclude with some final remarks in Sec.\ \ref{conclusion}. In the Appendix, we derive a new class of mirror trajectories associated with a Taub-NUT spacetime with vanishing 2-space curvature, $\varepsilon = 0$, where we discover a thermal spectrum at \textit{early-times}. We conjecture that the mirror trajectory and ensuing particle creation reflects black hole dynamics with a reversal of time, since in the $\varepsilon = 0$ case, time becomes a spacelike coordinate and vice versa. Throughout this paper, we utilise natural units, $G = c = \hslash = k_B = 1$.

\section{Accelerated Boundary Correspondence} \label{sec:metacc} 
\subsection{Taub-NUT metric}
\noindent The Taub-NUT metric  is often expressed in the form
\begin{align}\label{TNmet}
    \D s^2 &= - f (r) \big( \D \bar{t} - 2l \cos\theta \D \phi \big)^2 + \frac{\D r^2}{f(r)} + (r^2 + l^2 )\D \Theta^2,
\end{align}
where $\D \Theta^2 = \D \theta^2 + \sin^2\theta \D \phi^2$ and 
\begin{align}
    f(r) &= \frac{r^2 - 2Mr  - l^2}{r^2 + l^2}.
\end{align}
In the limit $l \to 0$, the metric reduces to the well-known Schwarzschild solution and $M$ takes on the interpretation of the mass of the source. The metric Eq.\ (\ref{TNmet}) has two Killing horizons but no curvature singularity. The physical meaning of the NUT parameter, $l$, remains an open question; it is commonly interpreted as a magnetic mass parameter \cite{Kerner:2006vu}, however other investigations have associated it with the twisting parameter of the source-free electromagnetic field within which a Schwarzschild black hole resides \cite{al2006physical}. The parameter $l$ relaxes global asymptotic flatness (e.g. singularity at $\theta = \pi$) despite the Riemann tensor scaling as $r^{-3}$ at infinity \cite{griffiths2009exact}.

Associated with the NUT charge $l$ 
is a singularity on the polar axis known as a Misner string (an analogue of the Dirac string singularity in electromagnetism) and regions of spacetime in its vicinity having closed timelike curves.  Traditionally these issues have been avoided by imposing the periodicity of the time coordinate \cite{Misner:1963fr},  rendering the string unobservable. This consequently leads not only to the existence of {closed} timelike curves everywhere, but additionally makes the maximal extension of the spacetime problematic \cite{Misner:1963fr,Hajicek1971CausalityIN}. However it has recently been shown that the spacetime described by the metric Eq.~\eqref{TNmet} is
geodesically complete and  free from causal pathologies for freely falling observers if time periodicity is not imposed \cite{Clement:2015cxa,Clement:2015aka}. Indeed, the Misner string is  transparent to geodesics, and 
the spacetime has no closed timelike or null geodesics provided some restrictions are imposed on the parameters of the NUT solution. Furthermore, the Kruskal extension through both horizons  can be carried out if there is no time periodicity \cite{Miller:1971em}, and it is possible to formulate a consistent  thermodynamics of Taub-NUT spacetime with Misner strings present \cite{Kubiznak:2019yiu,Ballon:2019uha,Bordo:2020kxm, Bordo:2019rhu}.

Because of the string singularity at $\theta = \pi$, one can perform the transformation $\bar{t} = t + 2l\phi$ to obtain the modified line element \cite{griffiths2009exact},
\begin{align}\label{TAUBNUT}
    \D s^2 &= - f (r ) \left( \D t + 4l\sin^2\frac{\theta}{2} \D \phi\right)^2 + \frac{\D r^2}{f(r) } + (r^2 + l^2 ) \D \Theta^2 .
\end{align}
There are two horizons, occurring at $r_\pm = M \pm \sqrt{M^2 + l^2}$, which define the so-called NUT regions, NUT$_-$, NUT$_+$. Consider the simplified (1+1)-dimensional metric in a plane of $\theta = \phi = \text{const.}$, yielding the simplified metric,
\begin{align}
    \D s^2 &= - f(r) \D t^2 + \frac{\D r^2}{f(r) }.
\end{align}
The thermal radiation emitted from the Taub-NUT black hole and detected by an inertial observer at infinity has temperature 
\begin{align}\label{temperature}
    T_\text{TN} &= \frac{\kappa_+}{2\pi},
\end{align}
where
\begin{align}\label{6}
    \kappa_+ &= \frac{1}{2} \frac{\D f(r)}{\D r}\bigg|_{r=r_+} = \frac{1}{2} \big( M + \sqrt{M^2 + l^2} \big)^{-1},
\end{align}
is the usual surface gravity at the outer horizon. From Eq.\ (\ref{temperature}), we see explicitly that the temperature of the black hole decreases with increasing $l$. This is reminiscent of the cooling effect that both charge and angular momentum have on the temperature of the RN and Kerr black holes respectively. We introduce a tortoise coordinate, obtained via
\begin{align}
     r^\star &= \int \frac{\D r}{f(r)},
\end{align}
which yields 
\begin{align}\label{tortoise}
    r^\star &= r + \sqrt{M^2 + l^2} \ln \bigg| \frac{r - r_+ }{r - r_-} \bigg| + M \ln \bigg| \frac{(r - r_+ )( r - r_- ) }{r_S^2} \bigg| ,
\end{align}
where $r_S \equiv 2M$ is the usual Schwarzschild radius, and an integration constant is chosen so that our results coincide with the Schwarzschild limit derived in \cite{Good:2016oey}, as $l \to 0 $. 

\subsection{Taub-NUT mirror}
\label{sec:K} 
\begin{figure}[h]
    \centering
    \includegraphics[width=\linewidth]{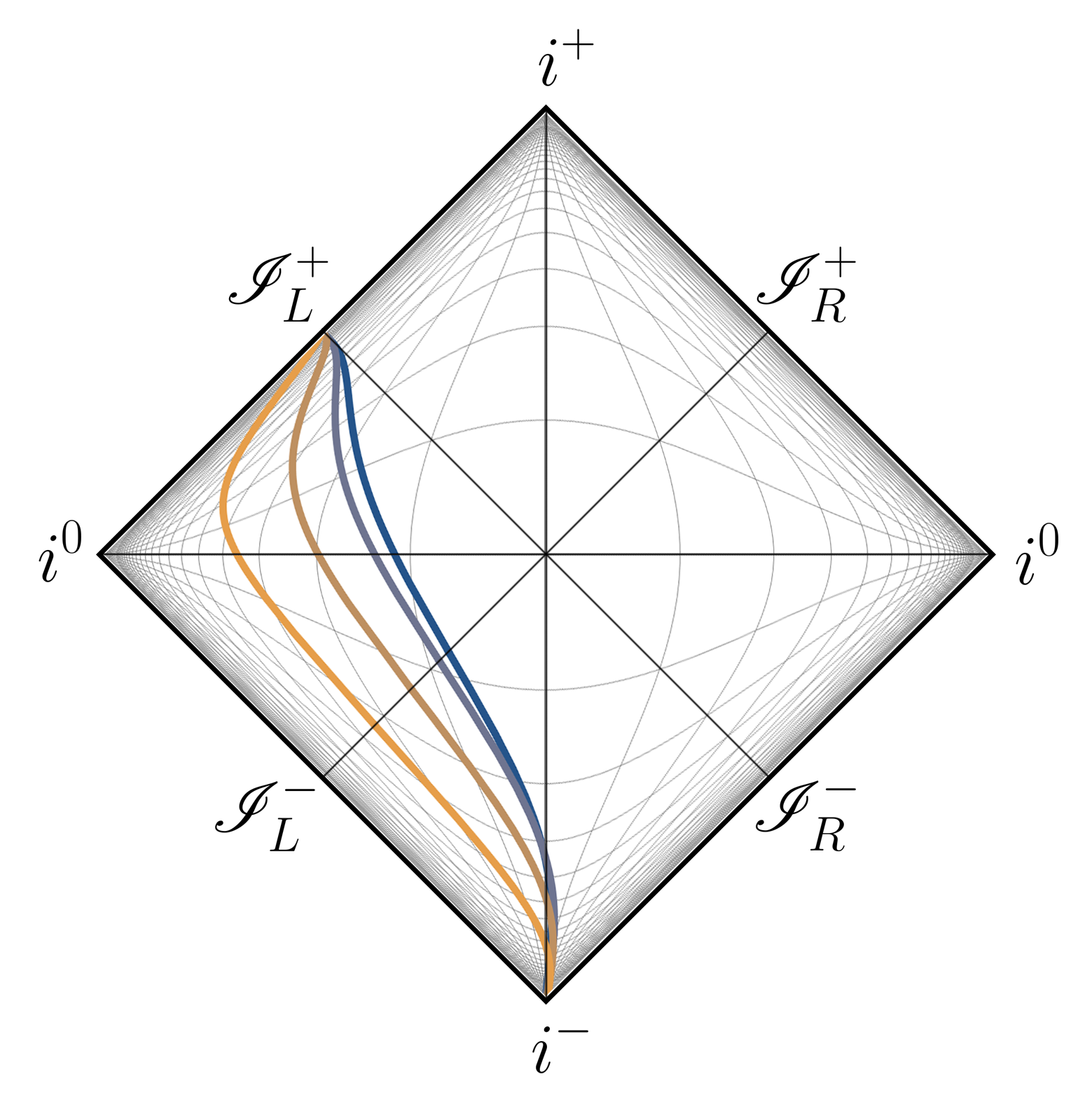}
    \caption{Penrose conformal diagram of the analog Taub-NUT mirror trajectories, with $l = 0.5$ and $M = 0.125$, $0.25$, $0.5$, $1$, ranging from dark blue to orange.  Recall that we have utilised natural units, setting $G = c = \hslash = k_B = 1$.} 
    \label{fig:penrose1}
\end{figure}

\begin{figure}
    \centering
    \includegraphics[width=0.75\linewidth]{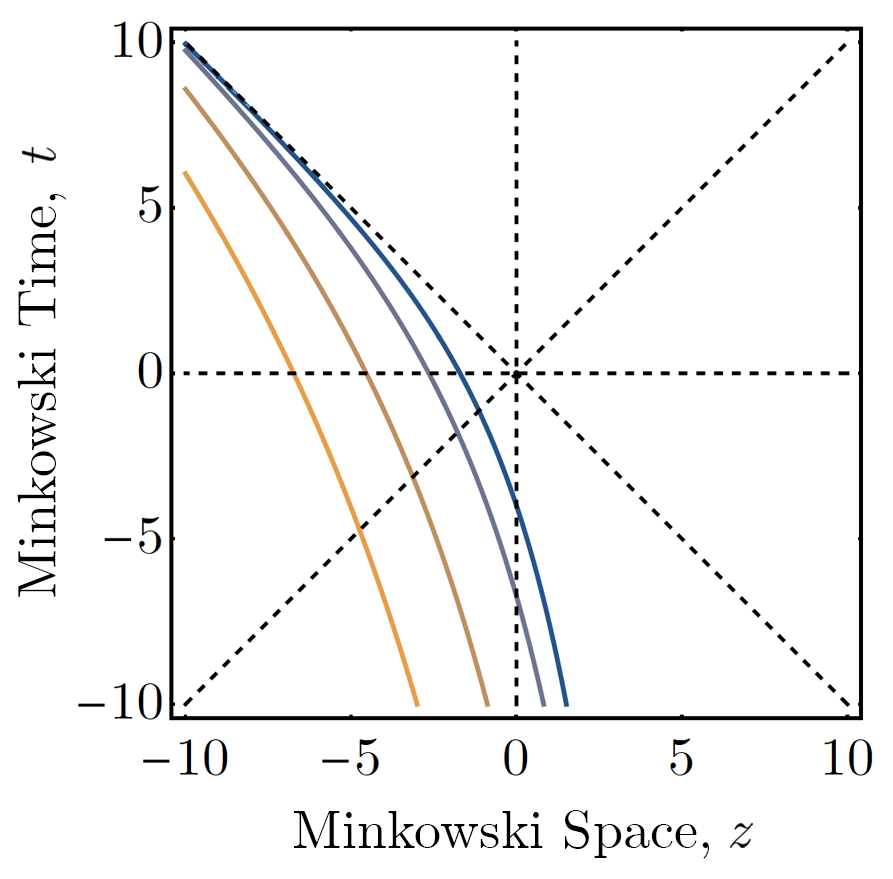}
    \caption{ Spacetime diagram of the trajectories shown in Fig.\ \ref{fig:penrose1}. These trajectories highlight the late-time acceleration horizon that mimics the formation of the black hole event horizon.}
    \label{fig:epsilononespacetime}
\end{figure}

The tortoise coordinate Eq.\ (\ref{tortoise}) can be used to defined a double null coordinate system, $(u,v)$, where $u = t - r^\star$ and $v = t + r^\star$, for the exterior geometry of the black hole, which allows the line element to be written in the form
\begin{align}
    \D s^2 &= -f \:\D u \D v.
\end{align}
One then employs a matching condition (see \cite{Fabbri}) with the flat interior geometry, which is described by the interior coordinates
\begin{align} 
    U &= T - r, \qquad V = T + r,
\end{align}
and associates this condition with the mirror trajectory, corresponding to the $r = 0$ coordinate. The matching condition expresses the exterior function, $u(U)$, in terms of the interior coordinate $U$. To do this, we set $r = r^\star$, and take $r^\star ( r = (v_0 - U) /2 ) = (v_0 - u )/2$ along a light ray, $v_0$. Using this condition, we obtain two possible conditions for $v_0$, namely $v_0 - 2r_\pm =  v_H$ where $v_H$ is the null coordinate of the horizon. Anticipating a transition to the (1+1)-dimensional mirror system, we can neglect the inner horizon solution, which occurs at $r<0$; the trajectory of the mirror (i.e.\ the reflecting point for incoming field modes) models the $r=0$ coordinate in the black hole coordinate system.  {More specifically, the particular choice of the light ray $v_0-2r_\pm = v_H$ gives us the analog to the outer horizon of the black hole. Since the field modes are lost to the left after $t=-x$ (the acceleration horizon of the mirror, see Fig.\ \ref{fig:penrose1} and \ref{fig:epsilononespacetime}), the inner horizon has a negligible role to play in determining the spectrum as seen by an observer on the right.} 

Applying these conditions to the tortoise coordinate, we obtain the exterior coordinate, $u(U)$, given by 
\begin{align}
    u(U) &= U - \frac{1}{\kappa_+} \ln \bigg| \frac{U}{4M} \bigg| -\frac{1}{\kappa_-} \ln \bigg| \frac{U -  4\sqrt{M^2 + l^2}}{4M} \bigg| ,
\end{align}
where
\begin{align}\label{11}
    \frac{1}{\kappa_\pm} &= 2M \pm 2\sqrt{M^2 + l^2},
\end{align}
are the inverse surface gravities of the outer and inner horizons respectively. 

As mentioned already, the accelerated boundary correspondence associates the origin of the black hole geometry with the trajectory of a perfectly reflecting point in (1+1)-dimensional Minkowski spacetime (i.e.\ our accelerated Dirichlet boundary condition). 
Making the identification $f(v) \Leftrightarrow u(U)$ (where $f(v)$ is commonly known as the ray-tracing function \cite{Birrell:1982ix}, i.e. it is the EoM trajectory of the mirror), we obtain 
\begin{align}
    f(v) &= v - \frac{1}{\kappa_+} \ln | \kappa_S v | - \frac{1}{\kappa_-} \ln | \kappa_S ( v - 4 \psi ) | ,
\end{align}
where $\kappa_S = 1/4M$ is the surface gravity of the Schwarzschild black hole and we have defined $\psi = \sqrt{M^2 + l^2}$. These black hole parameters retain their usual interpretation in the (1+1)-dimensional system, and their effect is to modify the spacetime trajectory of the mirror. It is straightforward to verify that this reduces to the Schwarzschild mirror trajectory in the limit $l \to 0$ \cite{Good:2016oey}. The rapidity $\eta(v)$ as a function of advanced time is given by $-2\eta(v)= \ln f'(v)$. For the Taub-NUT analog mirror, we obtain
\begin{align}
    \eta(v) &=  - \frac{1}{2} \ln \bigg| 1 - \frac{1}{\kappa_+ v}  + \frac{1}{4\kappa_- ( \psi  - v ) } \bigg| ,
\end{align}
which approaches the speed of light near the horizon, $v\to 0^-$. To leading order in $v$, the late-time proper acceleration, $\alpha(v) = e^{\eta(v)} \eta'(v)$, is given by 
\begin{align}
    \lim_{v\to 0^- } \alpha(v) &= - \frac{\kappa_+}{\sqrt{- 4 \kappa_+ v}}.
\end{align}
which is divergent. At early times, $v\to -\infty$, the mirror is static, as can be seen in the conformal Penrose diagram of Fig.\ \ref{fig:penrose1}.

\subsection{Energy flux and particle spectrum}
Having analysed the (1+1)-dimensional trajectory of the Taub-NUT analog mirror, we now consider the properties of outgoing particle and energy fluxes induced by its motion. The radiated energy flux, $F(v)$, can be calculated from the quantum stress-energy tensor using the simple expression \cite{Good:2016atu}, 
\begin{align}
    F(v) &= \frac{1}{24\pi} \big\{f(v) ,v \big\} f'(v)^{-2},
\end{align}
where the Schwarzian brackets are defined as
\begin{align}
\big\{f(v), v \big\} &= \frac{f'''}{f'} - \frac{3}{2} \left( \frac{f''}{f'} \right)^2 .
\end{align}
To leading order in $v$, near $v\to 0^-$, we discover a constant energy flux 
\begin{align}
    F(v) &= \frac{\kappa_+^2}{48\pi} + \mathcal{O}(v^2).
\end{align}
This behaviour is comparable to the analog mirror trajectories for the Kerr and Kerr-Newman black holes \cite{Good:2020fjz,Foo:2020bmv}, and is indicative of late-time thermal behaviour.

Next, we consider the particle spectrum of the outgoing modes. This can be derived from the Bogoliubov coefficients, 
\begin{align}\label{bogo}
    \beta_{\omega\omega' } &= \frac{1}{2\pi} \sqrt{\frac{\omega'}{\omega}} \int_{-\infty }^{v_H} \D v \:e^{-i\omega'v - i\omega f(v)},
\end{align}
where $\omega,\omega'$ are the frequencies of the outgoing and incoming modes respectively. This is a simplified form of the inner product integral in e.g.\ \cite{Good:2016atu} where integration by parts neglects non-contributing surface terms. The particle spectrum can be obtained by taking the modulus square, $\smash{N_{\omega\omega'}^\text{TN} := | \beta_{\omega\omega'} |^2}$ which yields
\begin{align}\label{TNparticle}
    N_{\omega\omega'}^\text{TN} &= \frac{\omega'}{2\pi \kappa_+ \omega_+^2} \frac{e^{-\pi\omega /\kappa_-}}{e^{2\pi\omega/\kappa_+} - 1} | U |^2 ,
\end{align}
where we have defined $\omega_+ = \omega + \omega'$, and
\begin{align}
    U \equiv U \left(  \frac{i\omega}{\kappa_-} ,  \frac{i\omega}{\kappa_S},  \frac{i\omega_+}{\bar{\kappa}} \right) ,
\end{align}
is a confluent hypergeometric Kummer function of the second kind, with  $\kappa_S =1/4M$  the usual surface gravity associated with  the Schwarzschild event horizon.
Here, $\bar{\kappa}^{-1} = 2(r_+-r_{-}) = 4\psi = 4\sqrt{M^2+l^2}$. This is analogous to the Kerr-Newman case with a replacement of the  angular momentum and charge with the NUT parameter (with a sign difference between them;  see Sec.\ \ref{KNTN} for a discussion of the rotating, charged scenario). 

\begin{figure}[h]
    \centering
    \includegraphics[width=\linewidth]{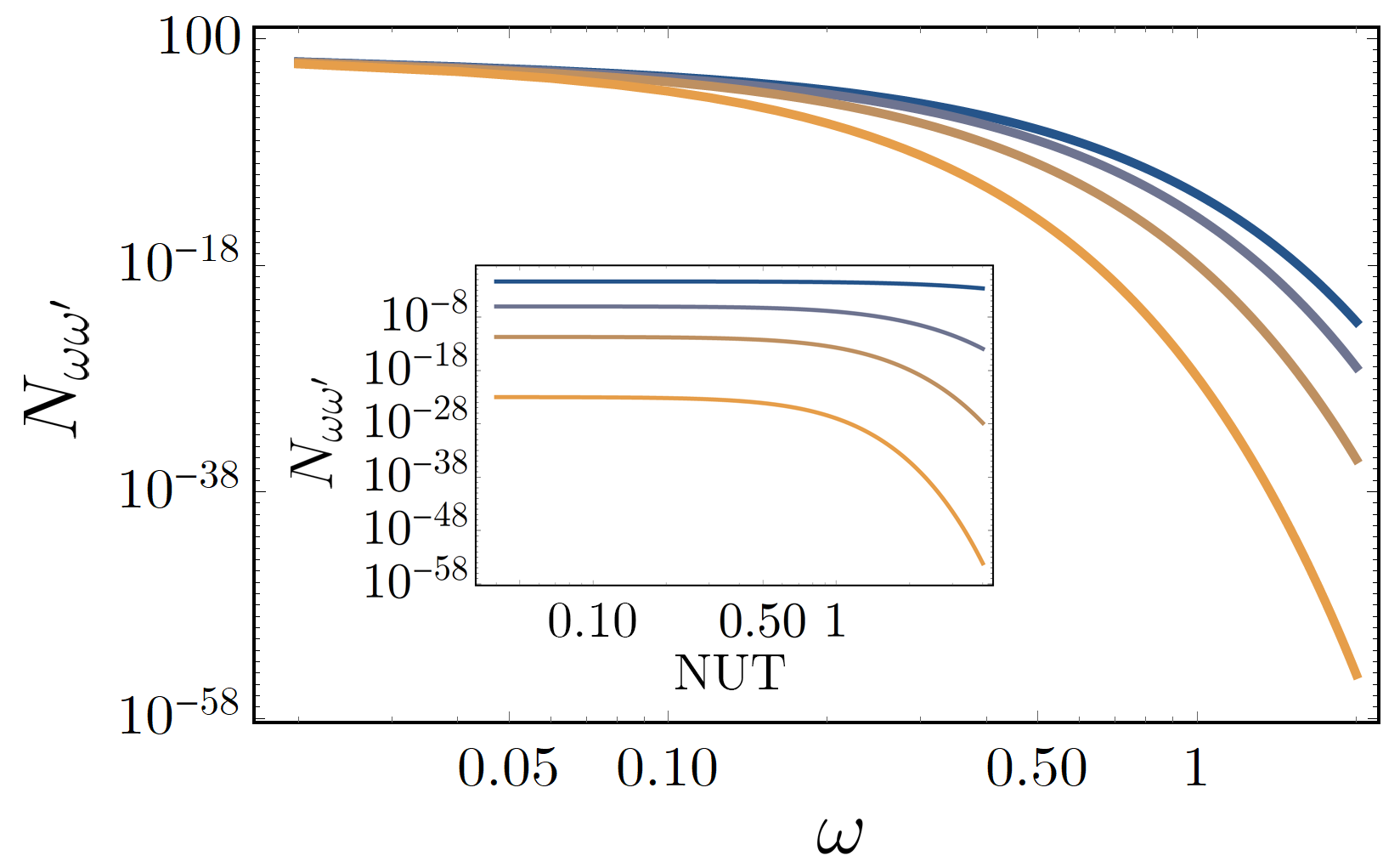}
    \caption{Mode-mode particle spectrum, Eq.\ (\ref{TNparticle}) emitted by the Taub-NUT (uncharged, non-rotating) analog mirror system, as a function of the outgoing frequency $\omega$ (main) and NUT parameter $l$ (inset). In the main plot, we have used $l = 0, 1, 2, 4$ (dark blue to orange) while in the inset, $\omega = 0.1$, $0.5$, $1$, $2$. In all cases, we have fixed $M = 1$ and $\omega'= 1$ for illustration.}
    \label{fig:modemode}
\end{figure}

As shown in Fig.\ \ref{fig:modemode}, $N_{\omega\omega'}$ approaches thermality in the late-time regime, associated with $\omega'\gg\omega$ \cite{Hawking:1974sw}. This limit describes the extreme Doppler shift experienced by the incoming plane wave modes, induced by the mirror trajectory. The main contribution to the Bogoliubov coefficients comes from these high-frequency modes. One can also demonstrate this thermal property analytically; that is,
\begin{align}
    \lim_{\omega'\gg \omega } N_{\omega\omega'}^\text{TN} &= N_{\omega\omega'}^\text{CW} = \frac{1}{2\pi \kappa_+ \omega'} \frac{1}{e^{2\pi\omega/\kappa_+} -1}, 
\end{align}
which is also the eternal Planckian spectral form obtained for the Carlitz-Willey mirror trajectory \cite{carlitz1987reflections} with temperature $T = \kappa_+/(2\pi)$ (i.e.\ associating $\kappa_+ \leftrightarrow \kappa$, where $\kappa$ is the analog surface gravity of the eternal black hole). 

It is also straightforward to verify that as the NUT parameter vanishes, the particle spectrum becomes
\begin{align}
    \lim_{l\to 0} N_{\omega\omega'}^\text{TN} = N_{\omega\omega'}^\text{S} &= \frac{\omega'}{2\pi \kappa_S \omega_+^2} \frac{1}{e^{2\pi\omega/\kappa_S}-1} ,\label{sch}
\end{align}
which is the result found in \cite{Good:2016oey} for the particle spectrum of the Schwarzschild analog mirror trajectory \cite{Good_2017Reflections,Anderson_2017,Good_2017BHII}.  {This limiting result, Eq.~(\ref{sch}), is valid for all-times, not just late-times, demonstrating the consistency of our approach with the canonical case.} 

From Fig.\ \ref{fig:modemode}, we also find that the introduction of the NUT parameter generally inhibits particle production, most clearly seen in the early-time limit (i.e.\ $\omega'\sim \omega$) in Fig.\ \ref{fig:modemode}. This is primarily due to the exponential suppression factor in Eq.~\eqref{TNparticle}, which even for small $l$ dramatically reduces $N_{\omega\omega'}$ by many orders of magnitude. From Eq.\ (\ref{11}), it can also be seen that the NUT parameter has a similar effect on the black hole surface gravity, and hence the temperature and particle production, as the mass (i.e.\ heavier black holes radiate fewer particles). Since the temperature has an inverse square-root dependence on $l$ -- Eq.\ (\ref{6}) -- the particle production is likewise suppressed for larger $l$ in the late-time thermal regime (however this effect is barely visible in Fig.\ \ref{fig:modemode}).

\section{Kerr-Newman Taub-NUT Mirror}\label{KNTN}
Thus far, we have considered a static, uncharged Taub-NUT spacetime and its analog mirror trajectory. Extension is warranted to the more general rotating, electrically charged Taub-NUT black hole. We note in passing that our analysis focuses on the pure (analog) Hawking radiation emitted by the mirror -- absent scattering effects \cite{Fabbri}. A limitation of the accelerated boundary correspondence is the neglect of higher-dimensional effects; in the rotating scenario, this includes superradiance, produced by incoming wave amplification due to scattering off the rotating black hole. However the influence of this effect primarily lies in the amplitude, rather than the frequency of the scattered modes. Hence in the following, we restrict our focus to the $s$-wave pure (analog) Hawking radiation emitted.

\subsection{Kerr-Newman Taub-NUT metric}
The Kerr-Newman Taub-NUT (KNTN) metric is given by 
\begin{align}\label{23}
    \D s^2 &= - \frac{\Delta }{\rho^2} \big( \D t - P \D \phi \big)^2 + \frac{\rho^2}{\Delta} \big( \D r^2 + \Delta \D \theta^2 \big) \nonumber \\
    & + \frac{\sin^2\theta}{\rho^2} \big( ( r^2 + a^2 + l^2 ) \D \phi^2  -a \D t \big)^2,
\end{align}
where
\begin{align}
    P &= a\sin^2\theta - 2l \cos\theta, \\ \label{25}
    \Delta &= r^2 - 2Mr + a^2 + Q^2 - l^2, \\ \label{26}
    \rho^2 &= r^2 + (l + a\cos\theta)^2.
\end{align}
Here, $a= J/M$ is the mass-normalised angular momentum and $Q$ is the charge.  {Following \cite{Good:2020fjz, Foo:2020bmv}, we further restrict our analysis to a plane of constant $\theta, \phi$
which yields the simplified (1+1)-dimensional metric
\begin{align}
    \D s^2 &= - f(r) \D t^2 + \frac{\D r^2}{f(r)},
\end{align}
where
\begin{align}\label{28}
    f(r) &= \frac{r^2 - 2Mr  + a^2 + Q^2 - l^2}{r^2 + (l+a\cos\theta)^2} .
\end{align}
From Eq.\ (\ref{28}), we find that the metric function, $f(r)$, is independent of $\phi$; hence the temperature of the Hawking radiation will likewise be unaffected by changes in $\phi$. However we also notice the presence of the angular coordinate $\theta$, which if left general, will leave an angular dependence in the temperature itself. To understand this, our model generates a correspondence between the (3+1)-dimensional black hole coordinates and the (1+1)-dimensional flat spacetime mirror trajectory by flattening out the two additional spatial dimensions defined by $\theta$, $\phi$. So far, this has been achieved simply assuming a plane of constant $\theta$, $\phi$ with arbitrary values, which has likewise yielded a valid tortoise coordinate which is \textit{independent of these parameters}. However the rotational degree of freedom in the full KNTN metric, Eq. (\ref{23}), leads to the existence of an ergosphere outside the black hole defined by $r_+ < r < r_{e+}$ where
\begin{align}\label{29}
    r_{e+} &= M + \sqrt{M^2 + l^2  - Q^2 - a^2 \cos^2\theta}.
\end{align}
Notice in particular that $r_{e+} = r_+$ (the outer horizon) when $\theta = 0$. In deriving the accelerated boundary correspondence between the black hole and the flat spacetime mirror trajectory, we require a tortoise coordinate defined with respect to the outer horizon, $r_+$, of the black hole. Thus, taking $\theta = 0$ yields a physically meaningful tortoise coordinate, and likewise, the correct outer horizon surface gravity. From this, we expect that the Hawking temperature of the outgoing radiation is the correct one, corresponding to that derived from other approaches, for example Eq.\ (\ref{6}). If $\theta \neq 0$, then one has an ill-defined tortoise coordinate which does not actually correspond to radial coordinate of the outer horizon.} 

With this in mind, we specialise to the plane of $\theta = 0$ and $\phi = \text{const.}$ which yields the tortoise coordinate
\begin{align}
    r^\star &= r + \frac{\gamma}{2\rho}  \ln \bigg | \frac{r-r_+}{r-r_-} \bigg| + M \ln \bigg| \frac{(r-r_-)(r-r_+)}{r_S^2} \bigg|,
\end{align}
where we have defined
\begin{align}
    \gamma &= 2al + 2l^2 + 2M^2 - Q^2, \\
    \rho &= \sqrt{M^2 + l^2 - a^2 - Q^2 }.
\end{align}
The spacetime also possesses two horizons at the radial coordinates 
\begin{align}
    r_\pm = M \pm \sqrt{M^2 + l^2 - a^2 - Q^2} .
\end{align}

\subsection{Kerr-Newman Taub-NUT mirror}
To obtain the exterior coordinate as a function of $U$, we perform the same matching condition analysis as before. The existence of two horizons for $r>0$ allows for the choice of $v_0  - 2r_+ \equiv v_H$ or $v_0 - 2r_- \equiv v_H$, since $u \to \infty$ at $U = v_H$. When $l^2<a^2+Q^2$, the inner horizon, $r_-$, occurs at a $r>0$, in contrast to the uncharged, non-rotating scenario. Without loss of generality, we set $v_H= 0$ and neglect the inner horizon solution so that $v_0 = 2r_+$. This choice is justified since it reduces to the correct Schwarzschild limit, wherein $r_- = 0$ represents the curvature singularity. Alternatively, one understands that the incoming modes are reflected off the center of the black hole coordinate system, $r = 0$. The outer radius is chosen for the shell position since $r_+>r_-$; the modes from
the shell $v_0=2r_+$ will reach the observer at $\mathscr{I}_R^+$ first in both the mirror and black hole system, having already reflected off the mirror.

The exterior coordinate is then,
\begin{align}
    u(U) &= U - \frac{1}{\kappa_+} \ln \bigg| \frac{U}{4M} \bigg| - \frac{1}{\kappa_-} \ln \bigg| \frac{U - 4\rho}{4M} \bigg|, 
\end{align}
where the inverse surface gravities of the inner and outer horizons are given by 
\begin{align}\label{34}
    \frac{1}{\kappa_\pm} &= 2M \pm  \frac{2al + 2l^2 + 2M^2 - Q^2}{\sqrt{M^2 + l^2 - a^2 - Q^2}}.
\end{align}
As before, we associate $u(U) \Leftrightarrow f(v)$ to obtain the trajectory of the mirror in (1+1)-dimensional Minkowski spacetime, given by
\begin{align}
    f(v) &= v - \frac{1}{\kappa_+} \ln \big| \kappa_S \nu \big| - \frac{1}{\kappa_-} \ln \big| \kappa_S (v - 4\rho ) \big|. 
\end{align}
The motion of the mirror is comparable to that found for the non-rotating, uncharged spacetime (see Fig.\ \ref{fig:penrose1} and \ref{fig:epsilononespacetime}); the parameters $M$, $Q$, $a$ only produce minor perturbations to the uncharged, non-rotating analog trajectory. 
The energy flux at late-times is given by 
\begin{align}
    F(v) &= \frac{\kappa_+^2}{48\pi} + \mathcal{O}(v^2) ,
\end{align}
which is constant, and dependent on $M$, $Q$, $a$, $l$. 
\begin{figure}
    \centering
    \includegraphics[width=0.99\linewidth]{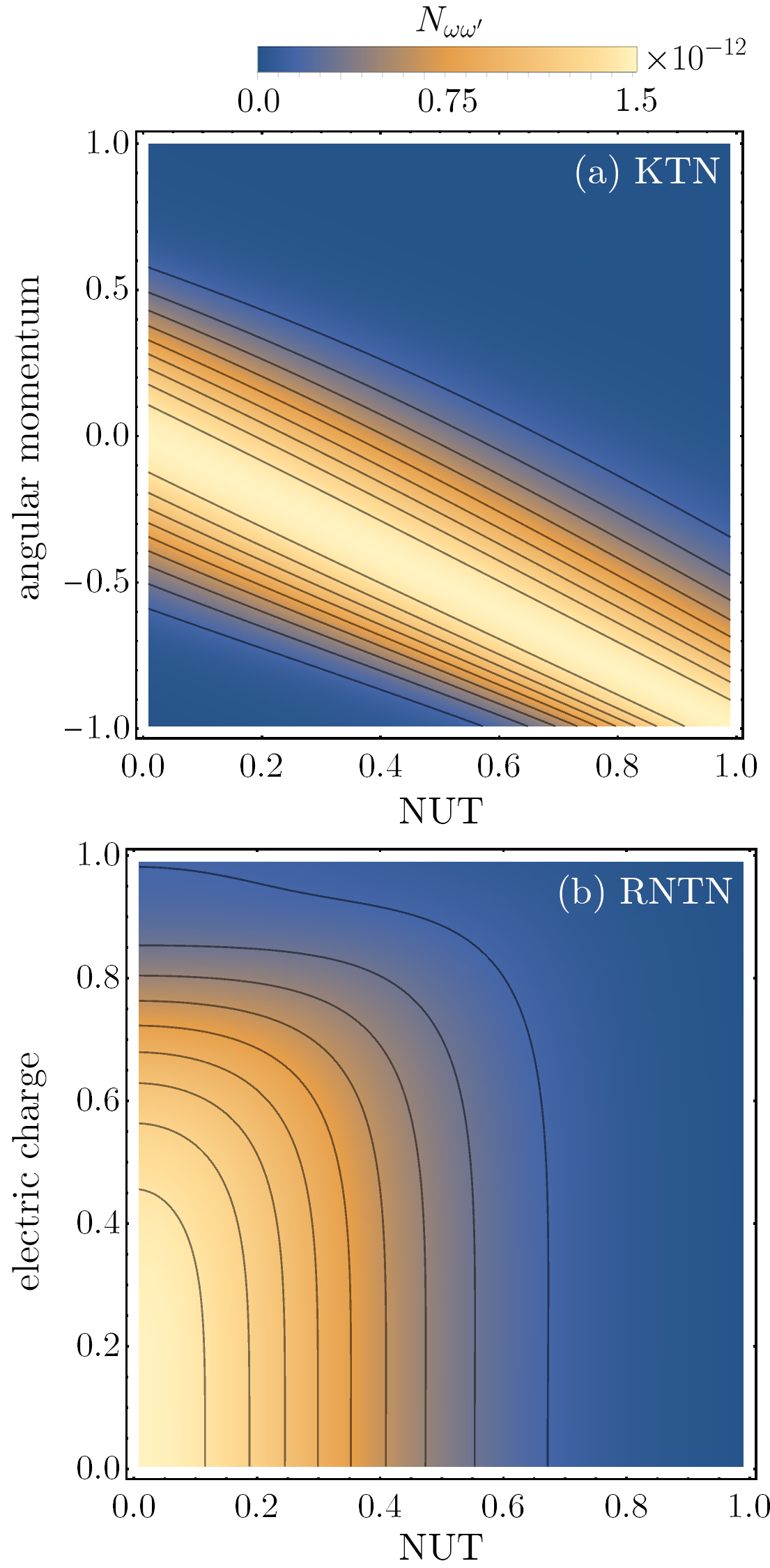}
    \caption{Mode-mode particle spectrum for the analog Kerr-Newman Taub-NUT mirror as a function (a) $a,l$ with $Q = 0$ and (b) $Q,l$, with $a = 0$. In all plots, we have assumed an early-time regime, $\omega\sim \omega'$ (i.e.\ $\omega = \omega' = 1$).}
    \label{fig:KNTNNww}
\end{figure}
As before, the mode-mode particle spectrum can be obtained via the Bogoliubov coefficients, and is given by
\begin{align}\label{KNparticle}
    N_{\omega\omega'}^\text{KNTN} &= \frac{\omega'}{2\pi \kappa_+ \omega_+^2} \frac{e^{-\pi\omega /\kappa_-}}{e^{2\pi\omega/\kappa_+} - 1} | U |^2 ,
\end{align}
where $\omega_+ = \omega + \omega'$ and again
\begin{align}
    U \equiv U \left(  \frac{i\omega}{\kappa_-} ,  \frac{i\omega}{\kappa_S},  \frac{i\omega_+}{\bar{\kappa}} \right) 
\end{align}
is a confluent hypergeometric Kummer function of the second kind, where $\kappa_S = 1/(4M)$ as before. Here, $\bar{\kappa}^{-1} = 2(r_+-r_{-}) = 4\sqrt{M^2+l^2-a^2-Q^2}$, which reduces straightforwardly to the Kerr-Newman case as $l \to 0 $, i.e.\ \cite{Foo:2020bmv}
\begin{align}
 \lim_{\omega'\gg\omega} \lim_{l\to 0}   N_{\omega\omega'}^\text{KNTN} = \lim_{\omega'\gg \omega} N_{\omega\omega'}^\text{KN} &= \frac{\omega'}{2\pi \kappa_+ \omega_+^2} \frac{ 1}{e^{2\pi\omega/\kappa_+} - 1}  
\end{align}
(where the surface gravities are those of the Kerr-Newman case). 

In general the spectrum is similar to the Taub-NUT case shown in Fig.\ \ref{fig:modemode} (hence, we leave out a comparable graph for the KNTN case for brevity). In the high-frequency limit, the outgoing particle flux is thermal, 
\begin{align}
    \lim_{\omega'\gg\omega} N_{\omega\omega'}^\text{KNTN} = N_{\omega\omega'}^\text{CW} &= \frac{1}{2\pi \kappa_+\omega'} \frac{1}{e^{2\pi\omega/\kappa_+}-1},
\end{align}
as expected. In the  $l^2 - a^2 - Q^2=0$ limit, the results reduce to the familiar Schwarzschild case, e.g.\ \cite{Good:2016oey}.
\begin{figure}[h]
    \centering
    \includegraphics[width=\linewidth]{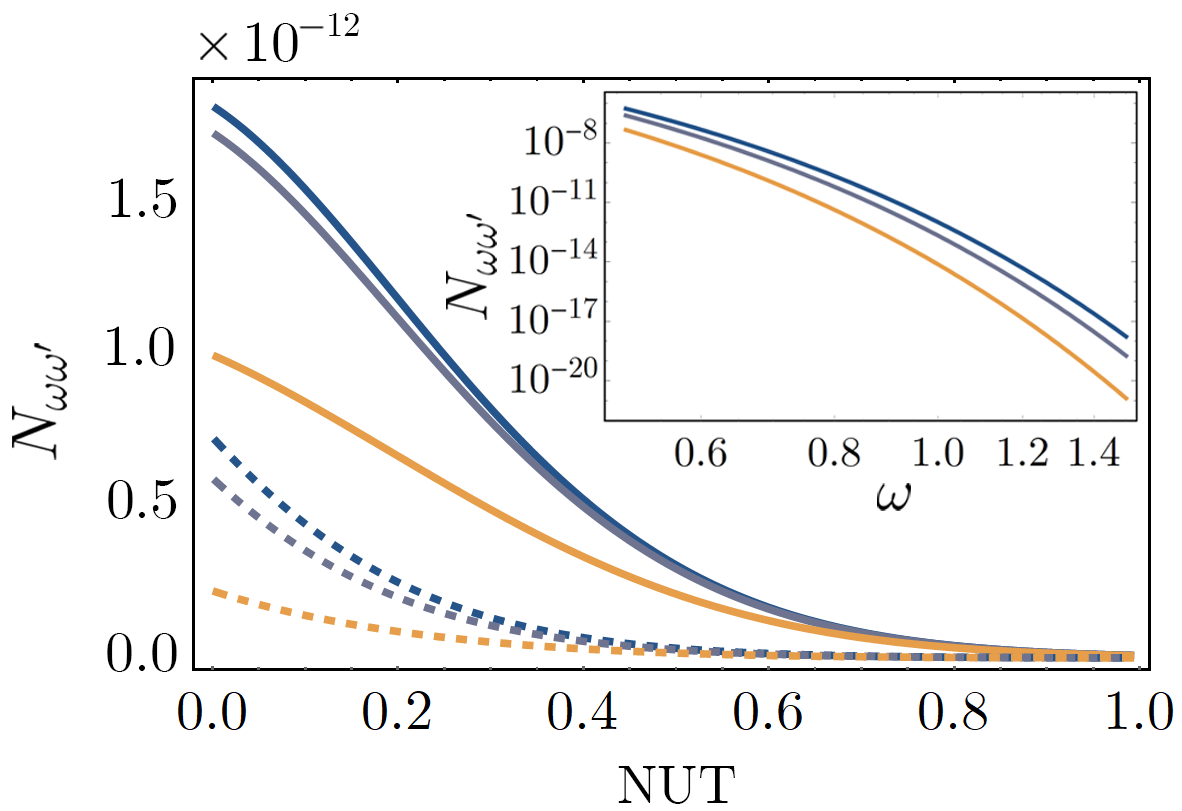}
    \caption{Mode-mode particle spectrum, Eq.\ (\ref{KNparticle}) for the Kerr-Newman Taub-NUT analog mirror. The main plot shows $N_{\omega\omega'}$ as a function of the NUT parameter, for different values of the charge; namely $Q = 0.1$, $0.4$, $0.7$ ranging from dark blue to orange, and the angular momentum, (solid) $a = 0.1$, and (dashed) $a = 0.7$. We have also taken $M = 1$ and the early-time limit, $\omega = \omega'= 1$. In the inset plot, $N_{\omega\omega'}$ is shown as a function of $\omega$, for $l = 0.0$, $0.5$, $1.0$ with $M = \omega'= 1$, $a = 0.1$, $Q = 0.7$ fixed. Note in particular that the inset plot is shown on a log-log scale for clarity.}
    \label{fig:KNTNl}
\end{figure}

\begin{figure}[h]
    \centering
    \includegraphics[width=\linewidth]{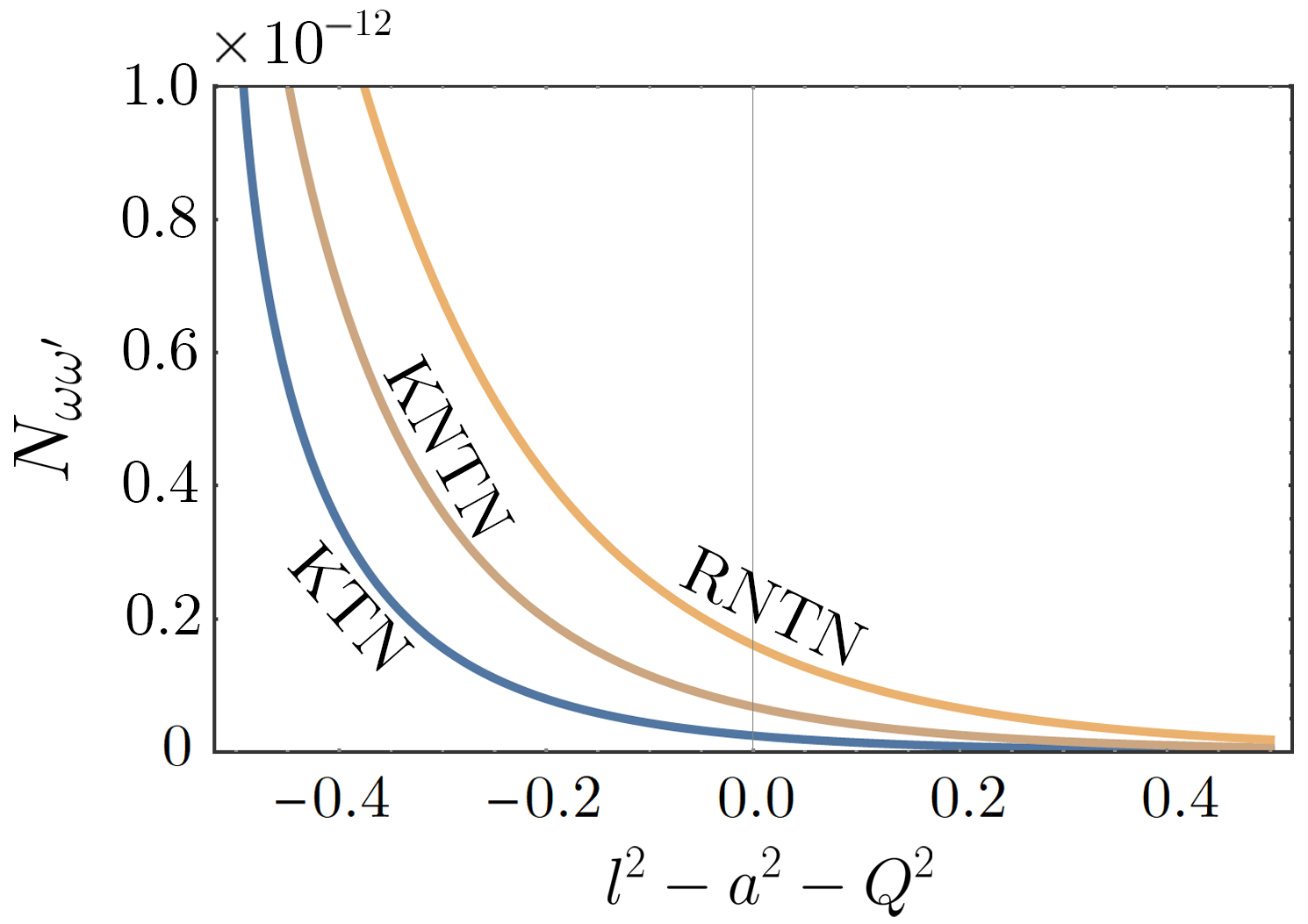}
    \caption{$N_{\omega\omega'}$ plotted against $l^2 - a^2 - Q^2$. The three lines correspond to (KNTN) $a = Q = 1/4$, (KTN) $a = 1/8$, $Q = 0$, and (RNTN) $a = 0$, $Q = 1/4$. We have also fixed $M = \omega = \omega' = 1$. Notably, $N_{\omega\omega'}$ exhibits no unusual behaviour as $l^2 - a^2 - Q^2$ crosses from negative to positive values. This behaviour occurs consistently across different incoming-outgoing frequency regimes.} 
    \label{fig:KNTNspectrum}
\end{figure}

To understand the dependence of particle production on the black hole parameters $(Q,a,l)$, it is instructive to plot the early-time spectrum, as shown in Fig.\ \ref{fig:KNTNNww}. This is because the early-time regime more explicitly unveils these dependences; the late-time regime yields the constant thermal production of the Carlitz-Willey trajectory. Figure \ref{fig:KNTNNww} shows the early-time spectrum for the (a) Kerr Taub-NUT and (b) Reissner-Nordstr\"om Taub-NUT analog mirrors. In Fig.\ \ref{fig:KNTNNww}(a), we find that for Kerr Taub-NUT black holes, the NUT parameter inhibits particle production, and at a faster rate than an equal increase in the black hole's rotation. Interestingly for Reissner-Nordstr\"om Taub-NUT black holes -- Fig.\ \ref{fig:KNTNNww}(b) -- the effect of the charge, $Q$, upon the radiated particle spectrum is nearly identical to that of the NUT parameter, $l$. That is, the early-time spectrum is nearly symmetric in $Q,l$. Further demonstration of the suppression of particle production by the presence of the NUT parameter is shown in Fig.\ \ref{fig:KNTNl}, which plots the early-time mode-mode spectrum of the Kerr-Newman Taub-NUT analog mirror. Figure \ref{fig:KNTNspectrum} shows that the particle spectrum does not exhibit any special behaviour as $l^2 - a^2 - Q^2$ crosses from negative to positive values. 

\section{Extremal Kerr-Newman Taub-NUT Mirror}
The extremal limit occurs for $M^2 = a^2 + Q^2 - l^2$, describing the minimum possible mass compatible with the other free parameters which characterise the Kerr-Newman Taub-NUT black hole.  Extremal black holes have been crucial in developing an understanding of the statistical origin of black hole entropy \cite{strominger1996microscopic}, making them relevant cases for studying quantum aspects of gravity.

\begin{figure}[h]
    \centering
    \includegraphics[width=\linewidth]{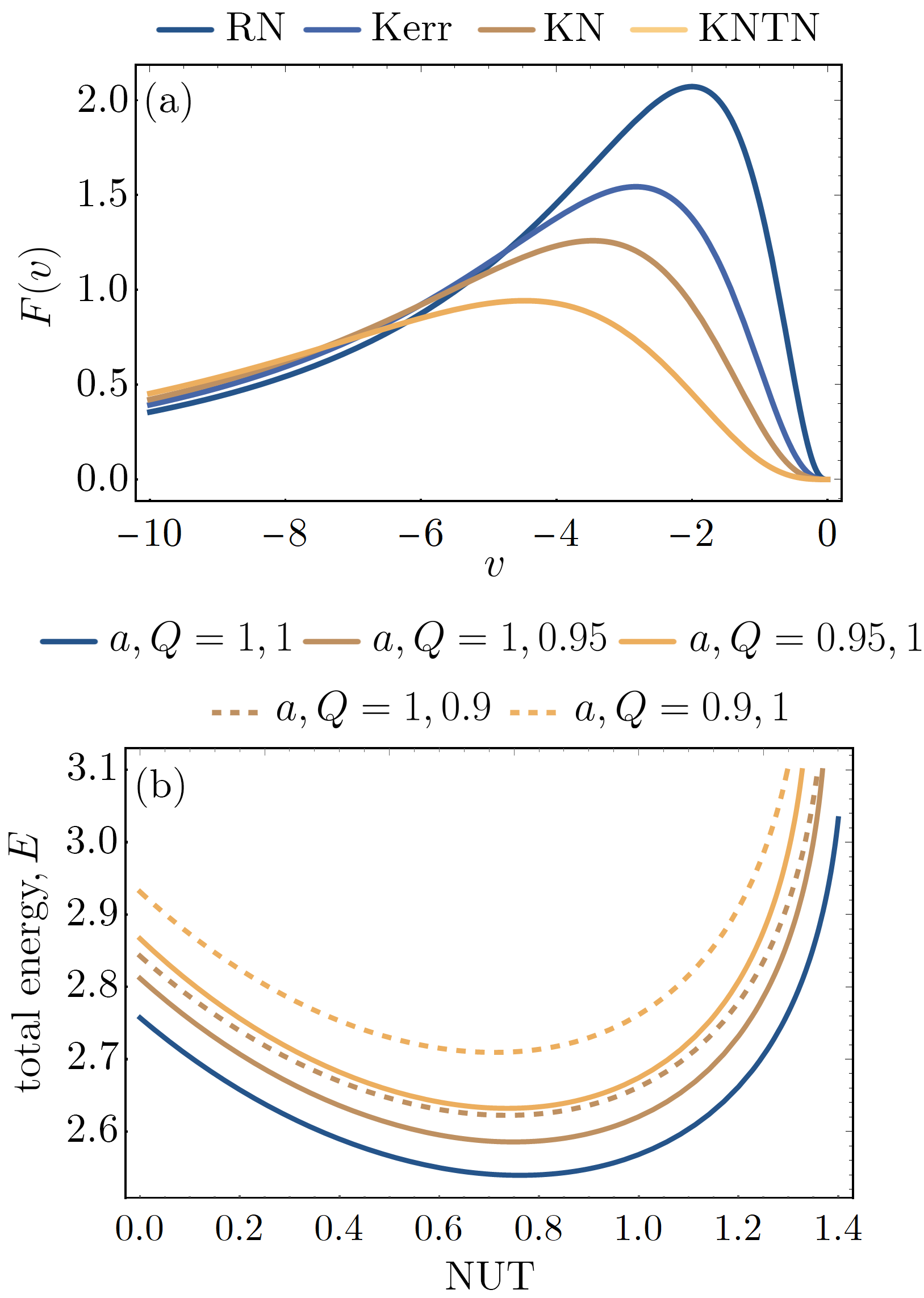}
    \caption{(a) Time-dependent energy flux, $F(v)$ (normalised by $10^{-4}$), plotted as a function of $v$ for the different extremal black hole solutions. The values of $a$, $Q$, $l$ are either $0$ or $+1$, corresponding to the relevant solution. (b) The total energy (normalised by $10^{-3}$) emitted by the KNTN black hole as a function of the NUT parameter.}
    \label{energyflux7}
\end{figure}

In this limit, the metric function becomes
\begin{align}
    f(r) &= \frac{r^2 - 2\sqrt{a^2 + Q^2 - l^2}r + a^2 + Q^2 - l^2}{r^2 + (l + a)^2} ,
\end{align}
taking the positive root of $M$. The tortoise coordinate is given by
\begin{align}\label{tortoiseextremal}
    r^\star &= r - \frac{2a ( a + l) + Q^2}{r- M } + 2M \ln \bigg| \frac{r- M }{2M } \bigg|.
\end{align}
In Eq.\ (\ref{tortoiseextremal}), we have restored the mass parameter $M = + \sqrt{a^2 + Q^2- l^2}$ where possible, bearing in mind that the extremal case is really characterised by three free parameters, rather than four. Performing the matching condition between the interior and exterior geometries of the black hole, we find that the exterior coordinate, as a function of $U$, is given by
\begin{align}
    u(U) &= U - \frac{4 (2a ( 2a + l ) + Q^2 ) }{U } - 4M \ln \bigg| \frac{U}{4M} \bigg|,
\end{align}
with the associated mirror trajectory given by 
\begin{align}
    f(v) &= v - \frac{1}{\mathcal{A}^2v} - \frac{1}{\kappa_S} \ln \big| \kappa_S v \big| ,
\end{align}
where $\mathcal{A}$ is defined in Eq.\ (\ref{mathcalA}). The rapidity is 
\begin{align}
    \eta(v) &= - \frac{1}{2} \ln \bigg| 1 + \frac{1}{\mathcal{A}^2 v^2} - \frac{1}{\kappa_S v} \bigg| ,
\end{align}
where we have anticipated the introduction of the asymptotic uniform acceleration, $\mathcal{A}$, defined as
\begin{align}\label{mathcalA}
    \lim_{v\to 0^- } \alpha(v) &= - \frac{1}{2 \sqrt{2a^2 + 2al + Q^2}} \equiv - \mathcal{A}.
\end{align}
Using the usual definition for the total energy flux, one obtains the following expression for the energy flux as a function of $v$,
\begin{align}
    F(v) &= \frac{\kappa_S^2\mathcal{A}^6 v^3 \left(\mathcal{A}^2 v (1-4 \kappa_S  v)+4 \kappa_S  (3 \kappa_S  v-1)\right)}{48\pi \left(\mathcal{A}^2 v (\kappa_S  v-1)+\kappa_S \right)^4}. \label{energyfluxext}
\end{align}
As was found for the extremal Kerr-Newman analog mirror trajectory, the energy flux emitted by the extremal Kerr-Newman Taub-NUT mirror, Fig.\ \ref{energyflux7}(a), vanishes at late times, $v\to 0$, and reduces to the result derived in \cite{Foo:2020bmv} as $l\to 0$.
The total energy radiated by the mirror is given by 
\begin{align}
    E &= \int_{-\infty }^{v_H=0} F(v) \frac{\D f(v)}{\D v}\D v \nn\\
 &= - \frac{\mathcal{A}\kappa_S}{48\pi \zeta^3} \bigg[ \mathcal{A}\zeta + \mu \bigg( \pi - 2 \tan^{-1} \left( \frac{\mathcal{A}}{\zeta} \right) \bigg) \bigg],\label{totenergyext}
\end{align}
with
\begin{align}\label{finiteenergy}
    \zeta &= \sqrt{4\kappa_S^2 - \mathcal{A}^2} , \\
    \mu &= \mathcal{A}^2 - 6\kappa_S^2 .
\end{align}
Analogous to other recent extremal black hole results (Kerr, Reissner-Nordstr\"om and Kerr-Newman mirrors), we find that the total energy radiated by the extremal Kerr-Newman Taub-NUT mirror is finite and reduces to the appropriate limits for a vanishing NUT parameter. 

In Fig.\ \ref{energyflux7}(b), we have plotted the total energy radiated by the mirror as a function of the NUT parameter. Intriguingly, for small $l$, the energy decreases, before increasing as $l \to \sqrt{a^2 + Q^2}$ -- that is, in the limit where the mass of the extremal black hole vanishes. It has been conjectured that tiny mass black holes were present in high densities in the early universe \cite{Zeldovich:1967lct}. The effect of this dominant energy contribution from primordial (i.e.\ tiny mass) Taub-NUT black holes upon the evolution of matter and energy densities in the early universe would be an interesting problem to pursue in future studies.
\begin{figure}[h]
    \centering
    \includegraphics[width=\linewidth]{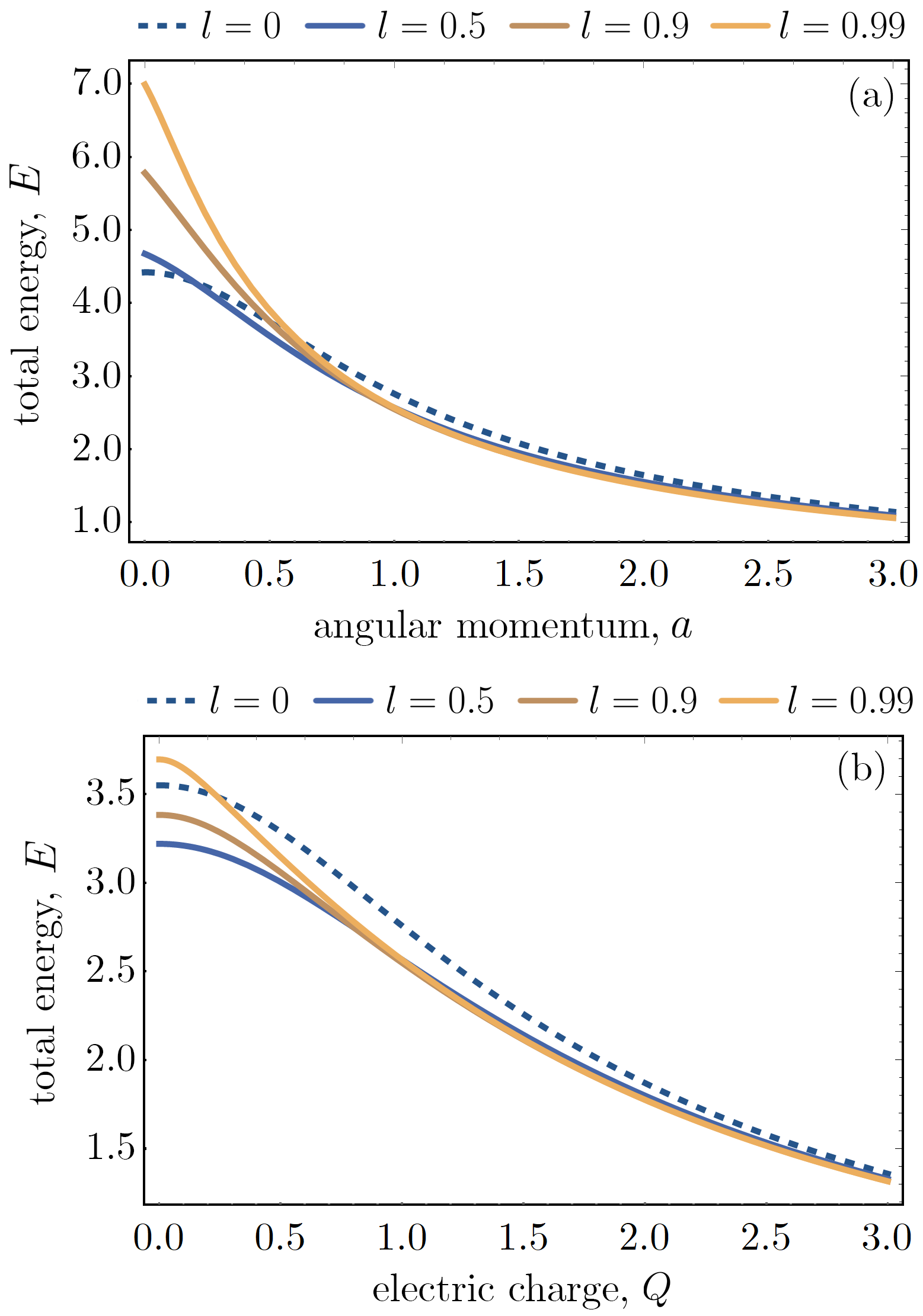}
    \caption{(a) Total energy, $E$, radiated by the extremal KNTN mirror as a function of the angular momentum for different values of the NUT parameter, with $Q = 1$ fixed. (b) Total energy, $E$, as a function of the charge, $Q$, with $a = 1$ fixed. Both plots have been normalised by $10^{-3}$. }
    \label{fig:energyvsA}
\end{figure}
In Fig.\ \ref{fig:energyvsA}(a), we have plotted the total energy emitted by the mirror as a function of the angular momentum parameter, $a$. In general, we see that the total energy decays with increasing $a$, which corresponds to an increasing extremal black hole mass. Notably, we find that for small values of $a$, the presence of a non-zero (and larger) NUT parameter amplifies the total energy. This result corroborates the findings of Fig.\ \ref{energyflux7}(b). At a threshold value of $a$, the energy of the $l=0$ mirror (i.e.\ the extremal Kerr-Newman analog) intersects that of the $l\neq 0$ mirrors; above this threshold, the presence of the NUT charge inhibits the total energy radiated compared with the $l = 0$ case. 

In Fig.\ \ref{fig:energyvsA}(b), we have plotted the total energy emitted by the mirror as a function of the charge, $Q$, with fixed $a$ and different values of $l$. The plot complements Fig.\ \ref{fig:energyvsA}(a), which shows that there is a specific value of $l$ for each set of $(Q,a)$ which minimizes the total energy radiated.

To derive the mode-mode particle spectrum, $N_{\omega\omega'}$, we perform the usual procedure and calculate the Bogoliubov coefficients between the incoming and outgoing modes. Performing this calculation yields, 
\begin{align}\label{extremalNww}
    N_{\omega\omega'}^\text{KNTN} &:= \big| \beta_{\omega\omega'}^\text{KNTN} \big|^2 = \frac{e^{-\pi\omega/\kappa_S}\omega'}{\pi^2 \mathcal{A}^2\omega_+} \left| K_n \left( \frac{2}{\mathcal{A}}\sqrt{\omega\omega_+} \right) \right|^2
\end{align}
where $K_n(x)$ is the modified Bessel function of the second kind, where $n= 1 - i \omega/\kappa_S$, and we have defined $\omega_+ = \omega + \omega'$. 
This spectrum is characteristically non-thermal, and accords with the results found in \cite{Good:2020byh,Foo:2020bmv} in the appropriate limits. It is important to note that the corresponding trajectory dynamics of the analog extremal mirror completely differ from the non-extremal case. This reflects the uniqueness of extremal black hole solutions, which possess vanishing surface gravity, and hence an undefined temperature. Likewise the non-thermal particle spectrum, Eq.\ (\ref{extremalNww}), should not be considered to be a limiting case of the non-extremal spectrum. 
\begin{figure}[h]
    \centering
    \includegraphics[width=\linewidth]{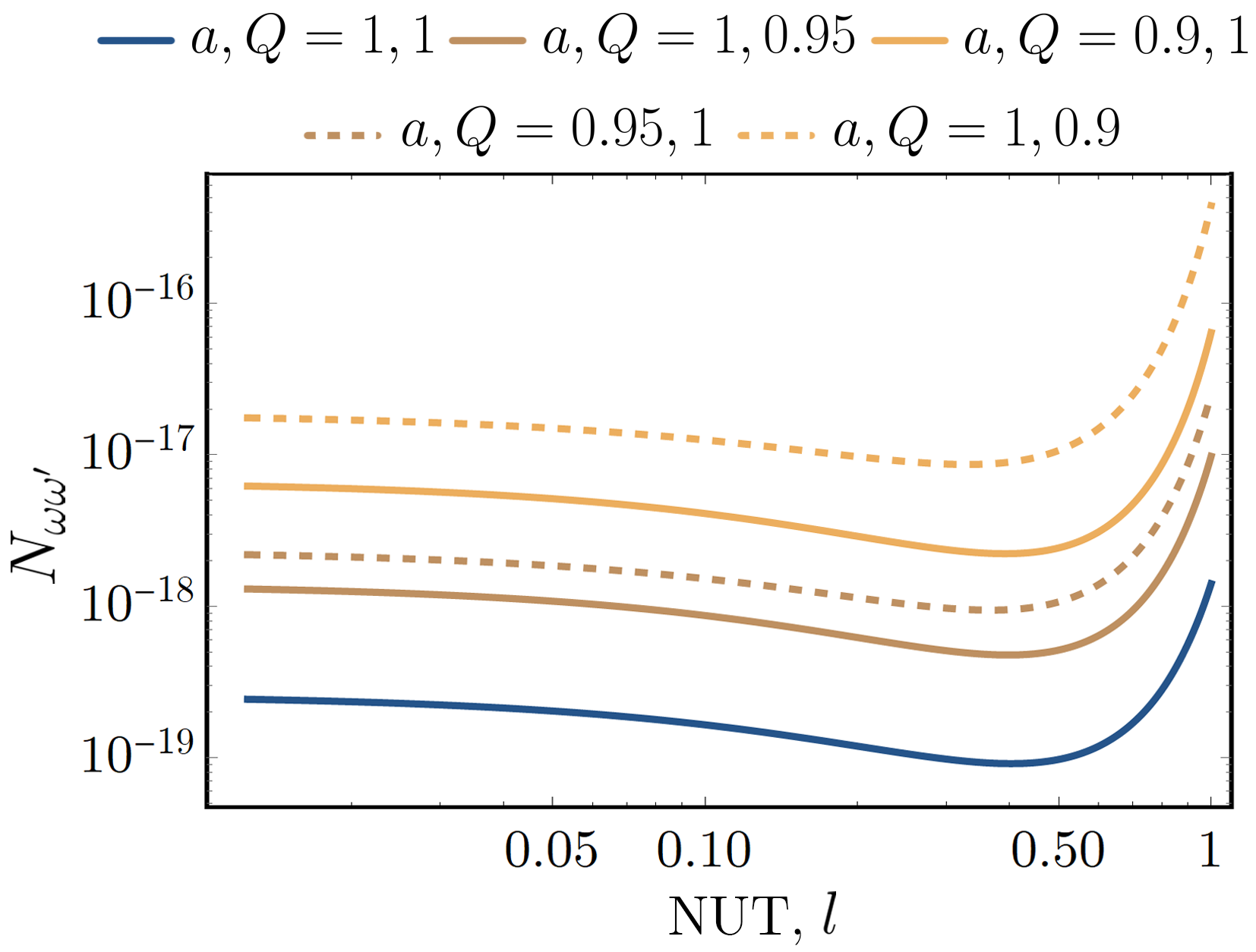}
    \caption{Log-log plot of the particle spectrum, $N_{\omega\omega'}$, radiated by the extremal KNTN mirror as a function of the NUT parameter, $l$. We have taken $\omega = \omega' = 1$, however we also verified that the non-monotonicity is robust to arbitrary choices of $\omega$, $\omega'$.}
    \label{fig:extremalnww}
\end{figure}
In Fig.\ \ref{fig:extremalnww}, we have plotted the early-time particle spectrum, $N_{\omega\omega'}$, for the extremal Kerr-Newman Taub-NUT analog mirror, as a function of the NUT parameter, $l$. We discover that the particle spectrum initially \textit{decreases} with increasing $l$ (i.e.\ as the mass of the extremal black hole \textit{decreases}). As the mass approaches zero, the particle number begins to increase and diverges as $l\to \sqrt{a^2 + Q^2}$. In the late-time regime, $\omega'\gg \omega$, the radiation exhibits similar behaviour. One can understand this as a competition between the increasing NUT parameter (which we have found for non-extremal black holes, inhibits particle production) with a decreasing black hole mass (which typically makes non-extremal black holes hotter; of course here, we cannot meaningfully assign a temperature to the radiation, since it is non-thermal). 
The particle spectrum Eq.~(\ref{extremalNww}) is in agreement with the energy flux of the quantum stress tensor, Eq.~(\ref{energyfluxext}). Numerical checks confirm that the method of quantum summing, 
\be E = \int_0^\infty \int_0^\infty \omega\; N_{\omega \omega'}^{\textrm{KNTN}} \d \omega \d\omega', \ee
yields the total energy Eq.~(\ref{totenergyext}), confirming the mathematical consistency of the spectral results. The physical interpretation is that the particles carry the energy; we conclude that the non-monotonic effect on the particle radiation as a function of the NUT parameter is a reliable result.

\section{Conclusion}\label{conclusion}

In this paper, we have investigated an accelerated boundary correspondence that mimics the outgoing Hawking radiation produced by a general class of Taub-NUT black holes. The solution is thermal at late-times, approaching the Schwarzschild and Carlitz-Willey limits in the appropriate regimes. In the rotating, electrically charged case, we find that the presence of the NUT parameter, $l$, generally suppresses particle production. Moreover, we have found no indication that the ABC form of the Hawking radiation spectrum responds to the type of singularity in the Taub-NUT metric that results in an absence of global asymptotic flatness. The extremal Kerr-Newman Taub-NUT case is a particularly interesting result, whereby the particle and energy spectrum are shown to be non-monotonic as a function of the NUT parameter, in contrast to the non-extremal case. 

As in other recent works, the aim of this paper has been to communicate the utility of the ABC method in extracting physically meaningful insights into the Hawking radiation properties of different kinds of black holes. This approach has the advantage of yielding analytic expressions for quantities like the particle spectrum and energy flux. We envision it will continue to be utilised as a tool for analysing complex and exotic cosmological spacetimes. 

\acknowledgments 

J.F.\ acknowledges support from the Australian Research Council Centre of Excellence for Quantum Computation and Communication Technology (Project
No.\ CE170100012). M.G. is funded from the state-targeted program ``Center of Excellence for Fundamental and Applied Physics" (BR05236454) by the Ministry of Education and Science of the Republic of Kazakhstan. M.G.\ is also funded by the FY2021-SGP-1-STMM Faculty Development Competitive Research Grant No. 021220FD3951 at Nazarbaev University.  R.B.M. acknowledges support from the Natural Sciences and Engineering Research Council of Canada.

\appendix

\section{Vanishing 2-Space Curvature ($\varepsilon = 0$)}\label{2space}
In our prior analysis, we have assume a NUT solution with positive 2-space curvature, $\varepsilon = 1$. More generally, the line element takes the form
\begin{align}\label{53metric}
    \D s^2 &= - f(r) \bigg[ \D t + \frac{il \big( \zeta \D \bar{\zeta} - \bar{\zeta} \D \zeta \big) }{1 + \varepsilon\zeta \bar{\zeta}/2} \bigg]^2 + \frac{\D r^2}{f(r) } \nonumber \\
    & + (r^2 + l^2 ) \frac{2 \D \zeta \D \bar{\zeta}}{(1 + \varepsilon\zeta \bar{\zeta} /2)^2},
\end{align}
where
\begin{align}
    f(r) &= \frac{\varepsilon(r^2 - l^2 ) -2Mr}{r^2  + l^2},
\end{align}
and $\varepsilon$ is the discrete, 2-space curvature which takes on the values $\varepsilon \in \{-1,0,1\}$. When $\varepsilon = + 1$, one can set $\smash{\zeta = \sqrt{2}\tan(\theta/2) e^{i\phi}}$ and the line element reduces to the form of Eq.\ (\ref{TAUBNUT}). In the following, we study the  
$\varepsilon = 0$ case, which corresponds to a new class of mirror trajectories possessing an early-time thermal spectrum.

\section{Taub-NUT Mirror ($\varepsilon = 0$)}
\noindent For $\varepsilon = 0$, the metric is 
\begin{align}\label{55metric}
    \D s^2 &= - f(r) \big( \D t + l\rho^2 \D \phi \big)^2 + \frac{\D r^2}{f(r)} + (r^2 + l^2) \D \Theta^2,
\end{align}
where
\begin{align}
    f(r) &= - \frac{2Mr}{r^2 + l^2}
\end{align}
and $\D \Theta^2 = \D \rho^2 + \rho^2 \D \phi^2$. To obtain Eq.\ (\ref{55metric}), we have set $\zeta = \rho e^{i\phi}/\sqrt{2}$ in Eq.\ (\ref{53metric}); in Eq.\ (\ref{53metric}), $\rho = 0$ behaves like an axis for which $\phi$ is the associated periodic coordinate \cite{griffiths2009exact}.  Conformal diagrams for sections of the maximally extended spaces with metric Eq.\ (\ref{55metric}) have been given by Siklos \cite{siklos1976two}. These correspond to the 2-space with $\rho, \phi = \text{const.}$, yielding the line element
\begin{align}\label{57metric}
    \D s^2 &= \frac{2M r}{r^2 + l^2} \D t^2 - \frac{r^2 + l^2}{2Mr} \D r^2. 
\end{align}
In this section, we will derive the accelerated boundary correspondence for Eq.\ (\ref{57metric}). Notice that the surfaces of constant $r$ have timelike normals for $r<0$ and spacelike normals for $r>0$. The usual ansatz assumed in the accelerated mirror model is that $r>0$, which accords with the regularity condition imposed on incoming modes. That is, the reflecting point of the modes is the $r = 0$ center of the black hole itself. Such a system (i.e.\ with timelike $r$) has not been studied in the context of accelerating mirrors, and hence the physical interpretation is not entirely clear. Curiously, if one makes the simple replacement $r\to - r$ (so that $r$ becomes the usual spacelike radial quantity), yielding the metric
\begin{align}
    \D s^2 &= - \frac{2Mr}{r^2 + l^2}\D t^2 + \frac{r^2 + l^2}{2Mr}\D r^2
\end{align}
the corresponding mirror trajectory becomes spacelike (admits faster-than-light trajectories), which is unphysical. 

\begin{figure}[h]
    \centering
    \includegraphics[width=\linewidth]{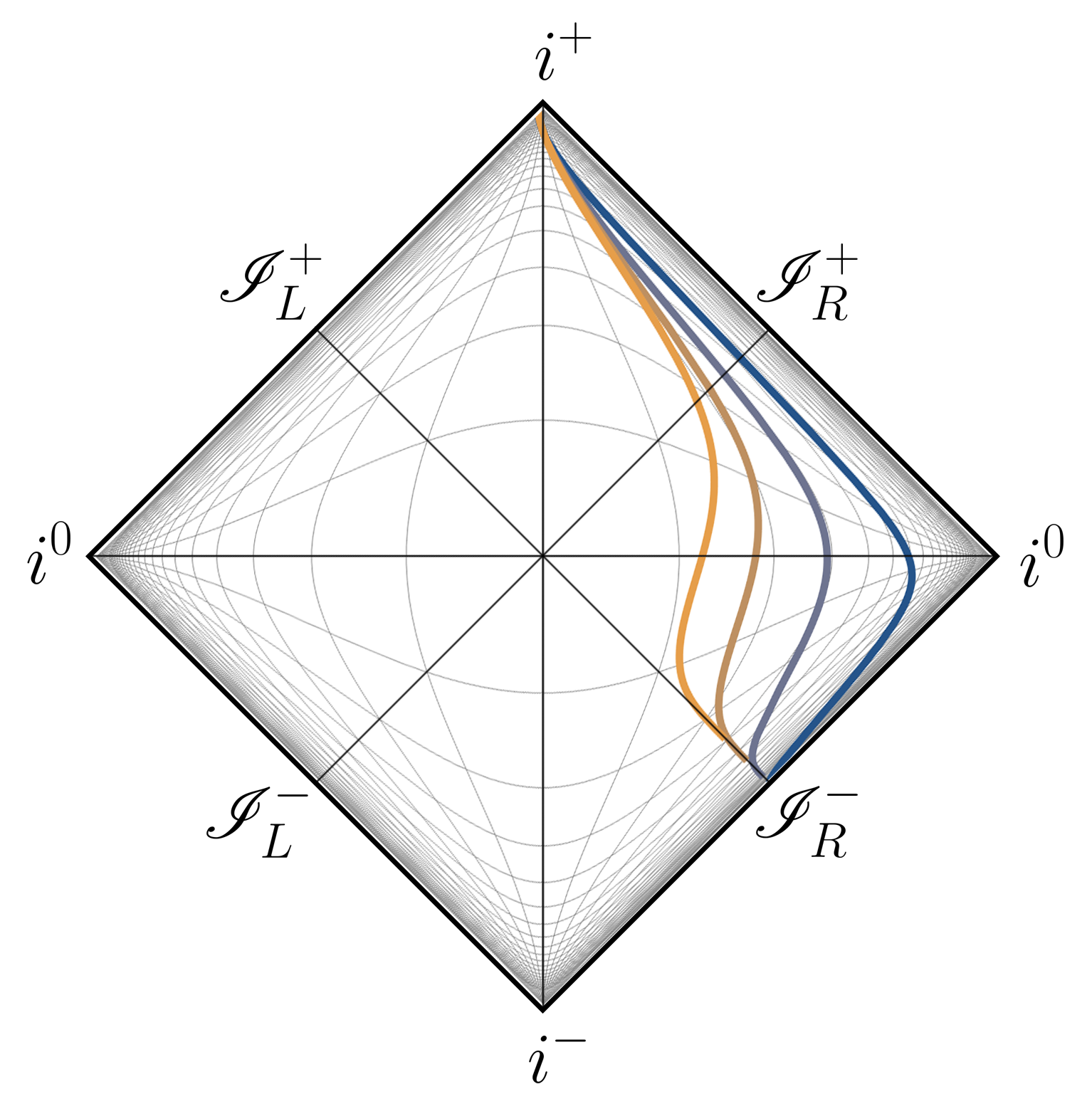}
    \caption{Conformal Penrose diagram for the analog Taub-NUT mirror trajectory, with vanishing 2-space curvature $\varepsilon = 0$. The trajectories shown correspond to $l = 0.5, 1.0, 1.5, 2.0$, from dark blue to orange.}
    \label{fig:zeroepsilon}
\end{figure}

\begin{figure}[h]
    \centering
    \includegraphics[width=0.85\linewidth]{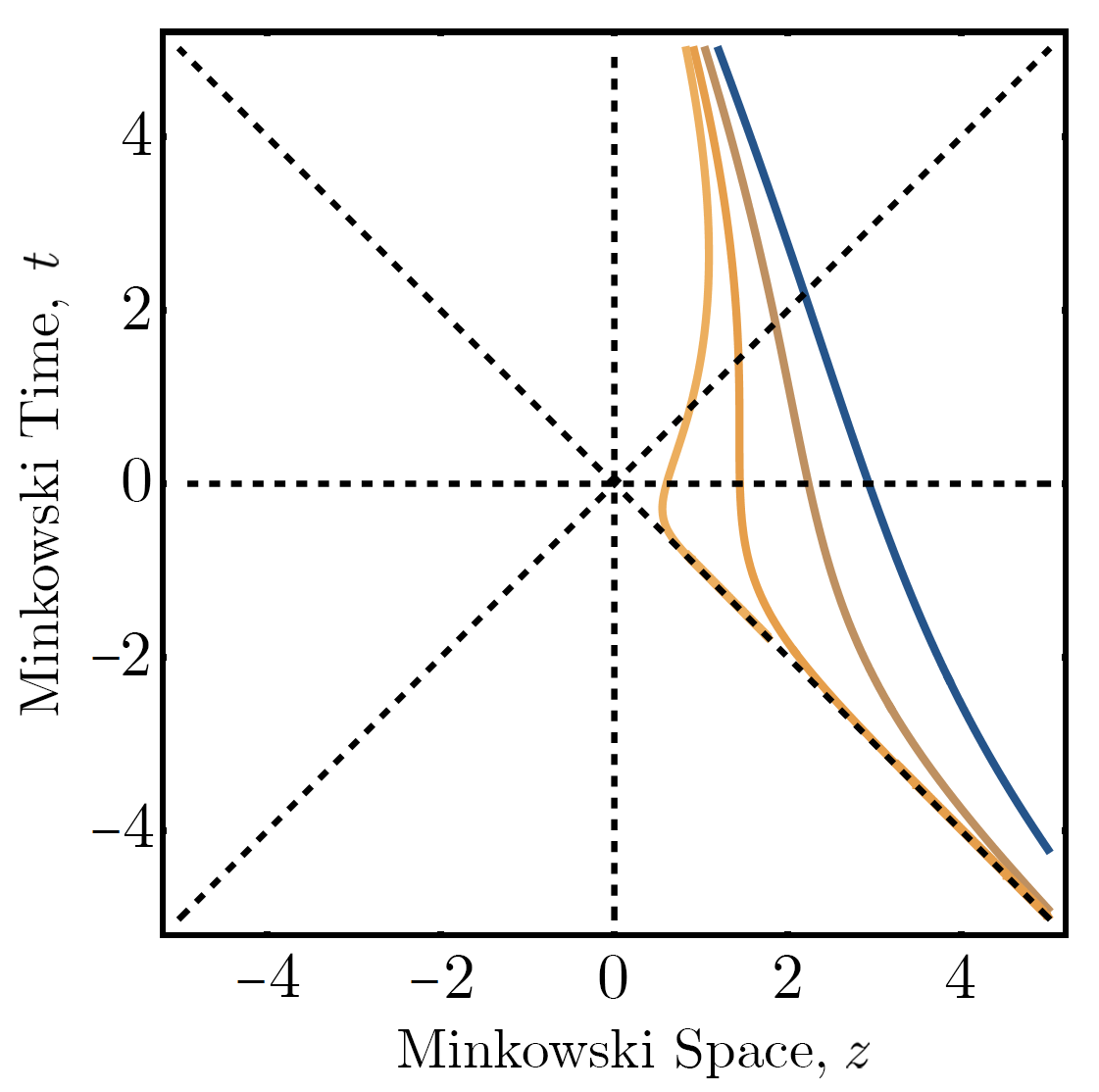}
    \caption{Corresponding spacetime trajectories for the Taub-NUT analog mirror, clearly illustrating the early-time horizon. We have used the same settings as Fig.\ \ref{fig:zeroepsilon}.}
    \label{fig:spacetimezero}
\end{figure}

Furthermore, adopting Eq.\ (\ref{57metric}) as stated leads to a valid spacetime trajectory possessing an \textit{early-time} acceleration horizon i.e.\ begins lightlike in the asymptotic past (see the trajectory diagrams in Fig.\ \ref{fig:zeroepsilon} and \ref{fig:spacetimezero}). Hence, we expect the spectrum to be thermal at early-times, rather than at late times as is usually the case for mirrors that approach the speed of light with an acceleration horizon in the asymptotic future. Tentatively, we suggest that the reversal in the sign of the time coordinate in the metric leads to an overall time reversal in the usual particle production dynamics of the mirror trajectory, leading to the early-time thermal result (as we derive below). Therefore, this mirror trajectory, and the ensuing spectrum, cannot have a direct physical correspondence to the behaviour of a typical black hole formed via gravitational collapse. Nevertheless, we are interested in the effects induced by this new class of mirror trajectories, which as we shall demonstrate, yield an entirely new particle and energy spectrum. Note that despite the swapped roles of $t$ and $r$ \textit{at the level of the metric \eqref{55metric}}, once we make the relevant associations between $u(U)$ and the flat spacetime trajectory $f(v)$, these details are flattened out; $t$ and $z$ take on their usual interpretations and temporal and spatial coordinates.

For the $\varepsilon = 0$ metric, the horizon occurs at $r = 0$ and is asymptotically flat as $r\to \pm \infty$. As mentioned, we consider $r\geq0$, where as usual $r = 0$ functions as the reflecting point of incoming modes. 
As before, we specialise to (1+1)-dimensions by considering a plane where $\theta = \phi = \text{const.}$, so that
\begin{align}
    \D s^2 &= - f(r) \D t^2 + \frac{\D r^2}{f(r)}.
\end{align}
The tortoise coordinate  (which is really a temporal tortoise coordinate \cite{Batic:2010vm}) is
\begin{align}
    r^\star &= - \frac{r^2}{4M} - \frac{l^2}{2M} \ln \bigg| \frac{r}{r_S^2} \bigg| ,
\end{align}
where we have an included an appropriately chosen integration constant in the denominator of the logarithm.  {Note that swapping $r \to - r$ at the level of the tortoise coordinate does not actually change $r^\star$; hence, the mirror trajectory remains identical. Furthermore, $r^\star \to +\infty$ as $r\to 0^+$}. As before, we apply the usual matching condition to the exterior $(u)$ and interior $(U)$ coordinates. Noting the single horizon at $r = 0$ and taking $v_0 = 0$ without loss of generality, we have
\begin{align}
    u(U) &= \frac{U^2}{8M} + \frac{l^2}{M} \ln \bigg| \frac{U}{2r_S^2} \bigg| .
\end{align}
The mirror trajectory can be obtained by associating $u(U) = f(v)$ so that
\begin{align}\label{epsilon0mirror}
    f(v) &= \frac{\kappa_S v^2}{2} + \frac{1}{\kappa} \ln \bigg| \frac{v}{2r_S^2} \bigg| 
\end{align}
where $\kappa_S$ is the usual Schwarzschild surface gravity and $\kappa = M/l^2$. We note here that the trajectory Eq.\ (\ref{epsilon0mirror}) represents a novel class of mirror trajectories that has not yet been studied. The rapidity is 
\begin{align}
    \eta(v) &= - \frac{1}{2} \ln \bigg| \frac{1}{\kappa v} + \kappa_S v \bigg| .
\end{align}
The mirror is asymptotically null in the infinite past and future, as shown in Fig.\ \ref{fig:zeroepsilon}.  In Fig.\ \ref{fig:spacetimezero}, we see that the trajectories diverge from null infinity, ${\cal I}_R^{-}$, and converge to timelike future infinity ${i}^{+}$. That is, even though both asymptotic regimes approach the speed of light, only the early-time regime possesses a null horizon. The late-time regime is asymptotically inertial.

One can also obtain the proper acceleration, which to leading order in $v$ near $v\to 0^-$ is 
\begin{align}
    \alpha(v) &= - \frac{\sqrt{\kappa_S v}}{2\kappa_S v^2} + \mathcal{O}(v^3) .
\end{align}
In the asymptotic past, the mirror possesses infinite acceleration (in the direction opposite to its motion) and coasts at the speed of light in the asymptotic future (and hence does not possess an acceleration horizon). As long as acceleration is asymptotically zero, even in the presence of a divergent rapidity (the mirror attains the speed of light), the mirror will be asymptotically inertial and its evolution will be toward asymptotic drift at constant velocity.  Coasting trajectories \cite{Good:2016atu,Good:2018ell,Good:2018zmx,Myrzakul:2018bhy,Good_2015BirthCry,Good:2016yht} have been studied as models for black hole remnants \cite{Chen:2014jwq}.  

\section{Energy flux and particle spectrum ($\varepsilon = 0$)}
The time-dependent energy flux, calculated using the usual Schwarzian derivative, is given by 
\begin{align}
    F(v) &= \frac{\kappa^2}{48\pi} \left[ \frac{1 + 10\kappa \kappa_S v^2 - 3\kappa^2 \kappa_S^2 v^4}{(1 + \kappa\kappa_S v^2)^4}\right] .
\end{align}
At early times $v\to 0^+$, the energy flux to leading order in $v$ is constant, and given by 
\begin{align}
    F(v) &= \frac{\kappa^2}{48\pi} + \mathcal{O}(v^2 ) \label{earlysteady}.
\end{align}
The constancy of $F(v)$ at early times is indicative of thermality, which is corroborated by the early-time spectrum for the particle production of the mirror,  discussed below. Meanwhile for late times, $v\to \infty$, the energy flux is negative and asymptotes to zero from below, 
\begin{align}
    F(v) &= - \frac{1}{16\pi \kappa_S^2 v^4} + \mathcal{O}(\lambda^5) . 
\end{align}
\begin{figure}[h]
    \centering
    \includegraphics[width=\linewidth]{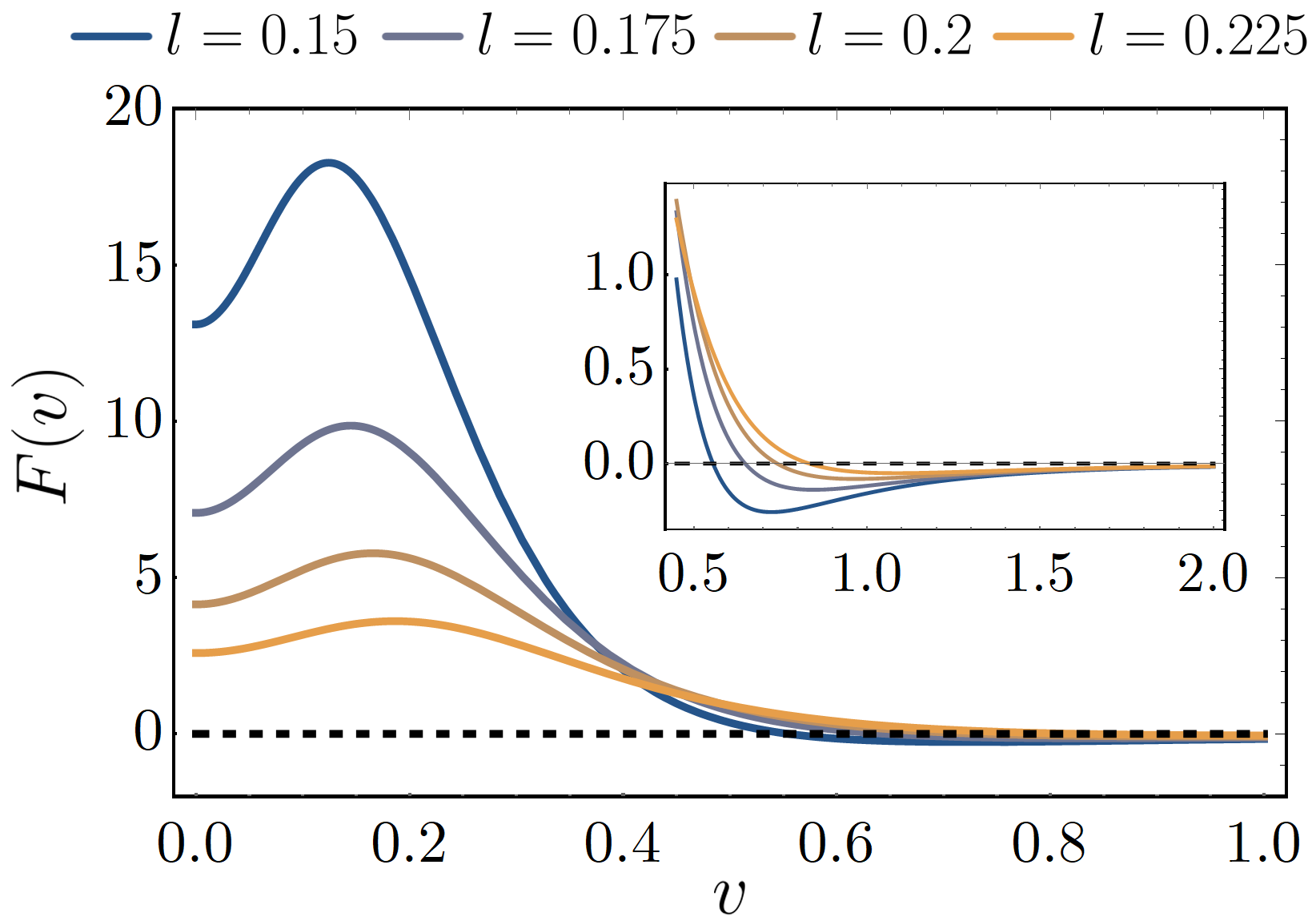}
    \caption{A plot of the energy flux as a function of the advanced time, with $M=1$ fixed. The inset shows the region of negative energy flux. }
    \label{fig:energyflux}
\end{figure}
Fig.\ \ref{fig:energyflux} displays the energy flux, $F(v)$, as a function of the advanced time coordinate $v$. The spectrum possesses two turning points at finite $v$, occurring at 
\begin{align}
    v_\pm &= \sqrt{\frac{3 \pm 2 \sqrt{2}}{\kappa \kappa_S}}.
\end{align}
After the initial burst of thermal particles, the flux increases towards a maximum at $v_+$, before decreasing and becoming negative and reaching a minimum at $v_-$. After this point, the energy emitted is negative into the asymptotic future. Negative energy emission from accelerated mirrors has been studied in \cite{Ford:1978qya,Davies:1982cn,Walker:1984ya,Ford:1990id,Ford:1999qv,Ford:2004ba}, and more pertinently, has been shown in settings where the trajectories are asymptotically coasting \cite{Good:2017kjr,Cong:2020nec}. Unitary evolution of conformal black holes evaporating non-monotonically require some transient period of negative energy flux \cite{Bianchi:2014qua,Good:2019tnf}. The presence of negative energy flux can also be understood in terms of outgoing modes whose quadratures are squeezed below the quantum shot-noise limit \cite{Su:2017fcm}. Note that as we have already mentioned, our results for the trajectory dynamics of the mirror and the corresponding particle and energy production make sense as a black hole analogy when time runs backwards. 

\begin{figure}[h]
    \centering
    \includegraphics[width=1\linewidth]{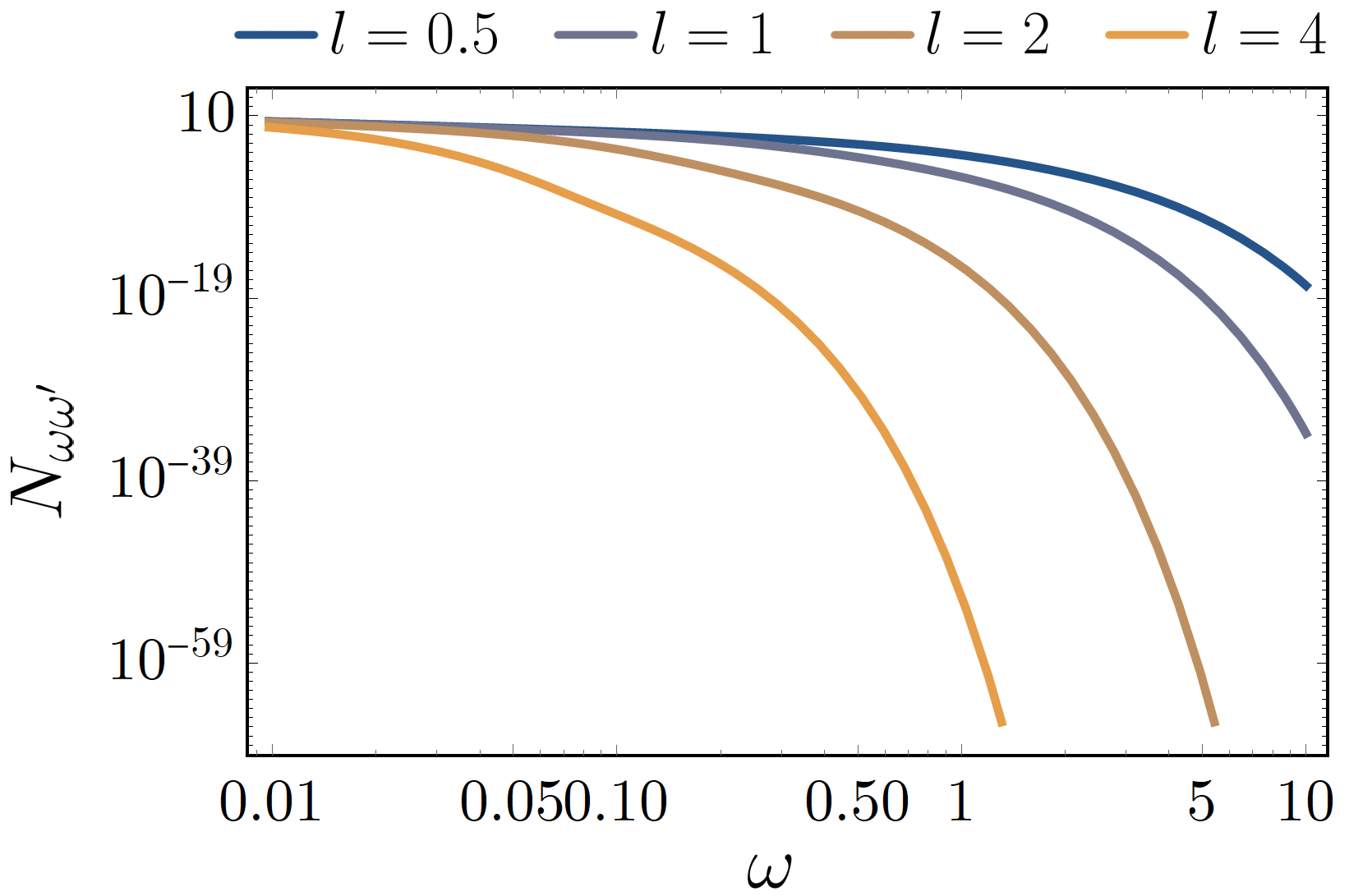}
    \caption{Mode-mode particle spectrum, Eq.~(\ref{f2}), for the Taub-NUT ($\varepsilon = 0$) analog mirror as a function of $\omega$ with $\omega'= 1$ fixed with $M= 1$ fixed. }
    \label{fig:epzerospectrum}
\end{figure}
We can also calculate the Bogoliubov coefficients, and hence the particle spectrum, analytically. The Bogoliubov coefficients are
\begin{align}
    \beta_{\omega\omega'} &= \frac{1}{2\pi} \sqrt{\frac{\omega'}{\omega}} \int_0^\infty \D v \:e^{-i\omega'v - i\omega f(v) }
\end{align}
noting that the integration domain is now $v\in [0,\infty)$. The particle number is given by 
\begin{align}
    N_{\omega\omega'}^{(\varepsilon = 0 )} := \big| \beta_{\omega\omega'} \big|^2 &= \frac{\omega' e^{-\pi\omega/2\kappa}}{8\pi^2\kappa_S^2 \omega^3}| \mathcal{F}|^2 \label{f2}
\end{align}
where $\mathcal{F} \equiv \mathcal{F}(\omega,\omega')$ is defined as 
\begin{align}
    \mathcal{F} &= i \sqrt{i\kappa_S \omega} \Gamma\left( \frac{\kappa-i\omega}{2\kappa} \right) \prescript{}{1}{F}_1 \left( \frac{\kappa-i\omega}{2\kappa} , \frac{1}{2} , \frac{i\omega^{\prime 2 } }{2\kappa_S \omega} \right) \nonumber \\
    & + \sqrt{2}\omega'\Gamma\left( 1 -\frac{i\omega}{2\kappa } \right) \prescript{}{1}{F}_1 \left( 1 - \frac{i\omega}{2\kappa} , \frac{3}{2} , \frac{i\omega^{\prime 2 } }{2\kappa_S \omega} \right)
\end{align}
and $\prescript{}{1}{F}_1(a,b,z)$ are confluent hypergeometric functions of the first kind, and $\Gamma(z)$ is the Gamma function. The result Eq.~(\ref{f2}) has radiated particles in a Planck distribution at early times to an observer on the right $\mathscr{I}^+_R$, 
\be \lim_{\omega'\gg\omega} N_{\omega\omega'}^{(\varepsilon = 0 )} = N_{\omega\omega'}^\text{CW} = \frac{1}{2\pi \kappa \omega'} \frac{1}{e^{2\pi \omega/\kappa}-1}, \label{f2planck}\ee
which can be found by the usual Hawking approximation \cite{Hawking:1974sw} $\omega'\gg \omega$ on Eq.~(\ref{f2}). This limit is complementary to the \textit{late-time} approximation taken for a mirror starting the past timelike infinity and receding to $\mathscr{I}^+_L$, for example, the $\varepsilon = +1$ trajectory. Since there is a steady-state energy flux emitted at early times, Eq.~(\ref{earlysteady}), one can see by using Eq.~(\ref{f2planck}) that the particles have temperature at $T = \kappa/(2\pi)$ at asymptotic early retarded times.
\begin{figure}[h]
    \centering
    \includegraphics[width=\linewidth]{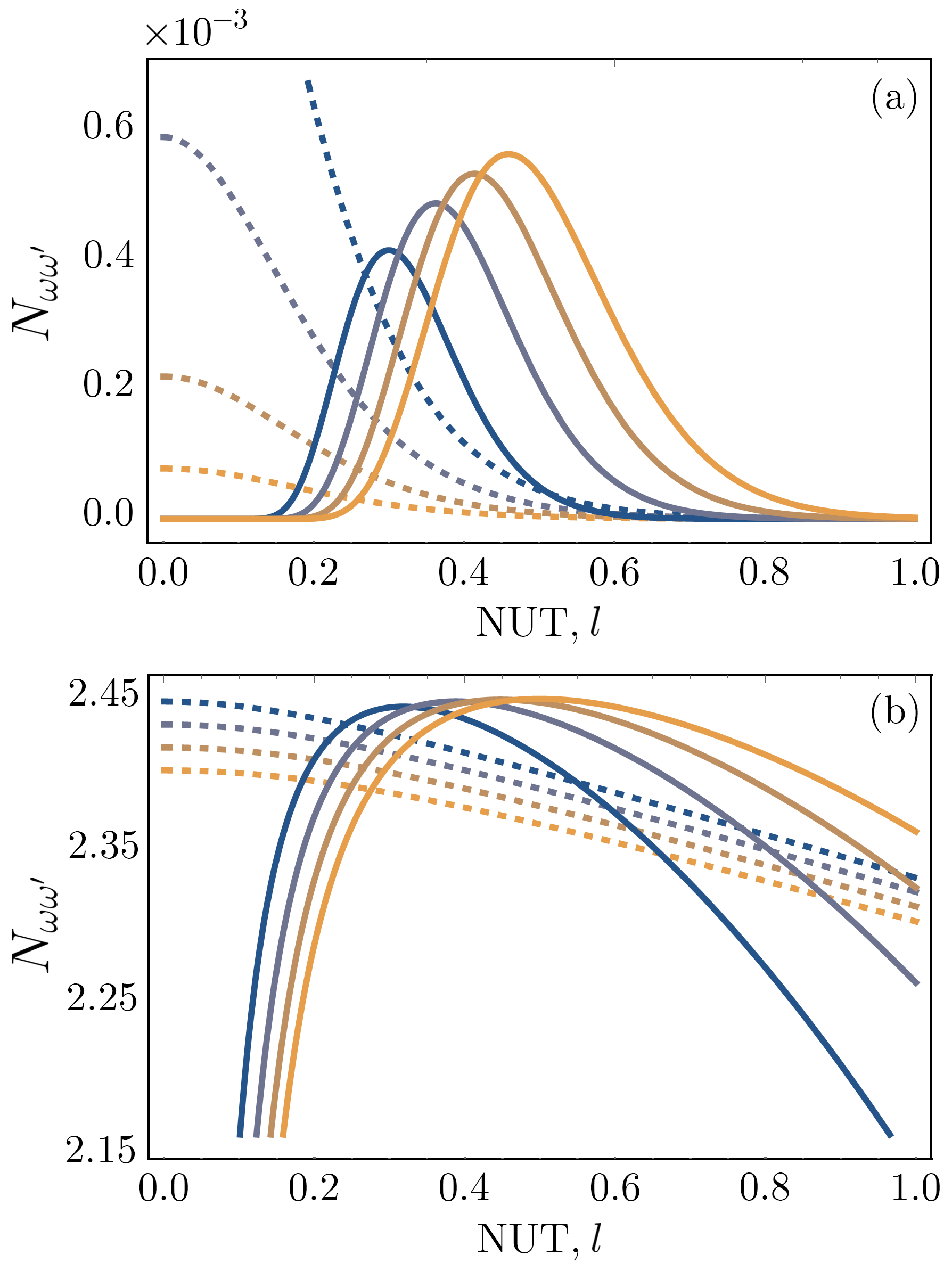}
    \caption{Plot of the particle number as a function of the NUT parameter, $l$, for different values of $M$. We have used (a) $\omega = 1$ and (b) $\omega = 0.01$, with $\omega' = 1$ in both plots. The dashed lines correspond to $N_{\omega\omega'}$ for the $\varepsilon = +1$ mirror, while the solid lines represent $N_{\omega\omega'}$ for the $\varepsilon = 0$ mirror. The colours (dark blue to orange) represent different black hole masses, $M = 0.1$, $0.15$, $0.2$, $0.25$ respectively.  }
    \label{fig:Nww01}
\end{figure}

Fig.\ \ref{fig:zeroepsilon} displays the mode-mode particle spectrum of the mirror for increasing (from dark blue to orange) values of the NUT parameter. From the values of $l$ for which we have plotted $N_{\omega\omega'}$, one might naively conclude that particle production is generally inhibited for larger NUT parameters. However we discover that $N_{\omega\omega'}$ is actually a non-monotonic function of the NUT parameter, as displayed in Fig.\ \ref{fig:Nww01}. The dashed lines correspond to the $\varepsilon = +1$ particle number, which decays monotonically with $l$, for all values of $M$, $\omega$, and $\omega'$. 

The solid lines, corresponding to $\varepsilon = 0$, show that the particle number vanishes in the limit of small $l$, a limit that was also verified analytically. For intermediate values of $l$, the particle number grows towards a maximum, the peak value of which depends on the mass of the black hole. Beyond this peak, the particle number decays to zero for large values of $l$. This behaviour is present for both the $\omega \sim \omega'$ and $\omega\ll \omega'$ regimes. 

The results shown in this Appendix have been motivated by the existence of a viable mirror trajectory obtained by the usual method from the metric of the $\varepsilon = 0$ Taub-NUT black hole. Our interest in this solution is primarily in the features of the derived mirror trajectory. The extent to which our result connects with the properties of a physical black hole remains an interesting direction for future research. Moreover, it would be interesting to study other systems in which $t$ and $r$ swap roles, and whether corresponding mirror trajectories generally give such behaviour.

\bibliography{main} 

\end{document}